# The Impact of Turbulence on Hydroacoustic Waves

Kai-Xin Hu[a,b,1], Yue-Jin Hu[c]

[a]Zhejiang Provincial Engineering Research Center for the Safety of Pressure Vessel and Pipeline, Ningbo University, Ningbo, Zhejiang 315211, China

[b]Key Laboratory of Impact and Safety Engineering (Ningbo University), Ministry of Education, Ningbo, Zhejiang 315211, China

[c]Ningbo Jiangbei District People's Government, Ningbo, Zhejiang, 315800, China

## Abstract

Traditional research suggests that when sound waves pass through a turbulent region, scattering occurs, causing the incident wave to attenuate and wave parameters to fluctuate. In contrast, our previous paper reported a new phenomenon in which turbulence causes changes in the amplitude of sound waves, a change that cannot be explained by scattering or resonance (HU, K.X, & HU, Y. J. 2025. Hydroacoustic Absorption and Amplification by Turbulence. arXiv:2512.07920). This work presents a more in-depth investigation into the impact of turbulence on hydroacoustic waves, including phase changes, amplification factors and the temporal evolution of the acoustic wave. Experiments indicate that turbulence simultaneously changes both the amplitude and phase of acoustic waves. The total phase shift along the entire pipe equals the sum of the phase shifts of the segments. Both the amplification factor and the phase shift due to turbulence vary periodically with frequency. In pipe flow, after the valve is closed, the temporal evolution of the acoustic waves during the subsequent turbulence decay process can be classified into six types. Acoustic waves with frequencies below and above specific thresholds are essentially unaffected by turbulence. In addition, vortices and unsteady flow in the laminar state do not cause changes in the amplitude and phase of sound waves, showing the essential difference between turbulent fluctuations and the two.

[1]Corresponding author, Email: hukaixin@nbu.edu.cn

## 1. Introduction

The influence of turbulence on acoustic waves is a fundamental problem shared by acoustics and fluid mechanics, and holds direct engineering significance in areas such as jet noise (Viswanathan 2024), cavity self-sustained oscillations (Rowley & Williams 2006), sonar detection (Komari & Farsi 2018), underwater communication (Li, Chitre & Stojanovic 2025), and atmospheric remote sensing (Killinger & Menyuk 2003). Understanding the effect of turbulence on sound is not only an inherent requirement of theoretical acoustics but also a practical necessity for improving the accuracy of acoustic detection, and has therefore long attracted the interest of researchers (Ostashev & Wilson 2026).

On the one hand, turbulence can act as a sound source to generate acoustic radiation. At low Mach numbers, Lighthill's acoustic analogy (Lighthill 1952) treats the Reynolds stress tensor in turbulence as an equivalent sound source. Building on this, Howe (2003) developed the theory of vortex sound, proposing that vortices radiate sound when they deform or annihilate during motion or interaction with boundaries, emphasizing vortex dynamics as the dominant mechanism of sound generation.

On the other hand, when sound waves propagate through a turbulent medium, their amplitude, phase, and coherence are all modulated by turbulent fluctuations. This understanding dates back to Obukhov's early investigations on the scattering and fluctuations of acoustic wave in the turbulent atmosphere (Kallistratova et al. 2018).

In atmospheric turbulence, when the acoustic wavelength lies within the inertial subrange of turbulence, the wavefront undergoes random distortion, giving rise to amplitude and phase fluctuations. Kallistratova's classic review (Kallistratova 2002) points out that for sound waves propagating in the the atmosphere, the scattering and fluctuations in the wave parameters are positively correlated with turbulence intensity and propagation distance. Random fluctuations in the temperature field cause local acceleration or deceleration of sound waves, leading to wavefront corrugation (Ostashev, Wilson & Colosi 2021). Meanwhile, velocity fluctuations modulate the apparent frequency of sound waves via the Doppler effect and induce scattering when the turbulence scale is comparable to the acoustic wavelength (Clair & Gabard, 2018).

In oceanic turbulence, inhomogeneities in the refractive index of seawater caused by turbulence lead to forward scattering and redistribution of acoustic energy, resulting in frequency-selective fading and pulse broadening effects (Tarng & Chang 1995). Bubble clouds generated by breaking waves at the sea surface are a significant source of low-frequency ambient ocean noise, with collective bubble oscillations radiating noise in the 30-500 Hz range (Carey & Fitzgerald 1993). Fluctuating pressures in the turbulent boundary layer can also excite structural vibrations, producing radiated noise (Ciappi et al. 2015). Furthermore, the disturbed flow field around an oscillating target causes scattering and modulation of acoustic waves, generating sideband harmonics in the frequency spectrum (Jing et al. 2025).

In terms of mechanisms, the influence of turbulence on acoustic waves, whether in the atmosphere or the ocean, can be understood within the framework of wave-turbulence scattering theory. The random inhomogeneity of the refractive index arises from turbulent fluctuations in temperature, salinity, or fluid velocity. Under the weak scattering approximation (i.e., turbulent fluctuations are much smaller than the mean sound speed), wave-turbulence scattering theory decomposes the sound field in such random media into a mean field and a fluctuating field. Using the Born approximation or the Rytov method to solve the wave equation, the theory predicts the variances of the fluctuations in sound pressure amplitude and phase, coherence loss, and spectral broadening (Ostashev, Wilson & Colosi 2021; Ostashev & Wilson 2026).

In the study of the interaction between sound waves and turbulence, classical scattering theory typically assumes that turbulence is unaffected by sound waves and that acoustic energy merely undergoes spatial redistribution. However, this assumption holds only when the sound wave frequency is much higher than the characteristic frequency of turbulence. When their time scales are comparable, sound waves can be absorbed by turbulence through resonance.

Ingard & Singha (1974) found in experiments on gas in pipes that the attenuation mainly originates from viscous dissipation and heat conduction near the pipe walls. Howe's (1984) theoretical study showed that attenuation is caused by the strain generated by the sound field near solid walls, and is maximized when the time scales of sound waves and turbulence are matched. Subsequent experiments and numerical studies by Peters et al. (1993) and Weng, Boij & Hanifi (2013, 2015) further validated the above conclusions.

Different from the scattering, resonance, viscous dissipation, etc., of sound waves in turbulence in the aforementioned studies, we have discovered in underwater acoustic experiments a new phenomenon of amplitude variation of sound waves under the influence of turbulence (Hu & Hu 2025). The flow conditions included both pipe flow and free jet flow, while the frequency range spanned from 60 kHz to 4.4 MHz. This study found that acoustic waves can be significantly absorbed and amplified by turbulence, even at frequencies far exceeding the turbulent fluctuation frequencies, with no observed spectral broadening or additional frequency components.

In pipe flow, the total amplification factor of acoustic waves in the entire pipeline is equal to the product of the amplification factors in each section of the pipeline. The amplification factor depends on the wave frequency rather than its amplitude. Turbulent fluctuations without mean motion can still alter the wave amplitude, while laminar flow has no effect on acoustic signals. Similar phenomena were observed in jet turbulence, with no evidence of wave scattering by turbulence detected.

These findings suggest that the primary cause of changes in acoustic wave amplitude due to turbulence is the turbulent fluctuations themselves, rather than the mean flow or scattering. Comparisons with conventional theories and experimental studies indicate that the observed phenomena in this study are not attributable to bubbles, resonance, scattering, or viscous dissipation. This indicates that there is a new mechanism in the interaction between turbulence and acoustic waves that has not yet been fully elucidated, requiring further experiments to uncover it.

This study will further explore the influence of turbulence on hydroacoustic waves based on the previous paper, including phase changes, the evolution of amplitude over time, and their dependence on frequency, among other aspects. In Section 2, we present the experimental setup, and in Section 3, the experimental results are listed. We demonstrate how turbulence alters the phase of waves, how the amplification factor varies with frequency, the types of temporal evolution observed in received signals, and experimental results for low and high frequency waves. The effects of standing waves are discussed, and the differences in how vortices, unsteady flow, and turbulent fluctuations influence sound waves are compared. Finally, Section 4 provides a summary.

## 2. Experimental Setup

The experimental setup in this study is shown in Figure 1. Two hydroacoustic transducers, located at both ends of the pipeline, are connected to a signal generator and an oscilloscope, respectively, serving as the acoustic wave transmitter and receiver. In the experiment, a single-frequency sinusoidal signal is generated by the signal generator. The external water flow is driven by a water pump, entering from the pipeline inlet and exiting from the outlet. When propagating along the pipeline axis, the acoustic waves are affected by turbulence. The amplitude and phase of the received signal are observed through an oscilloscope.

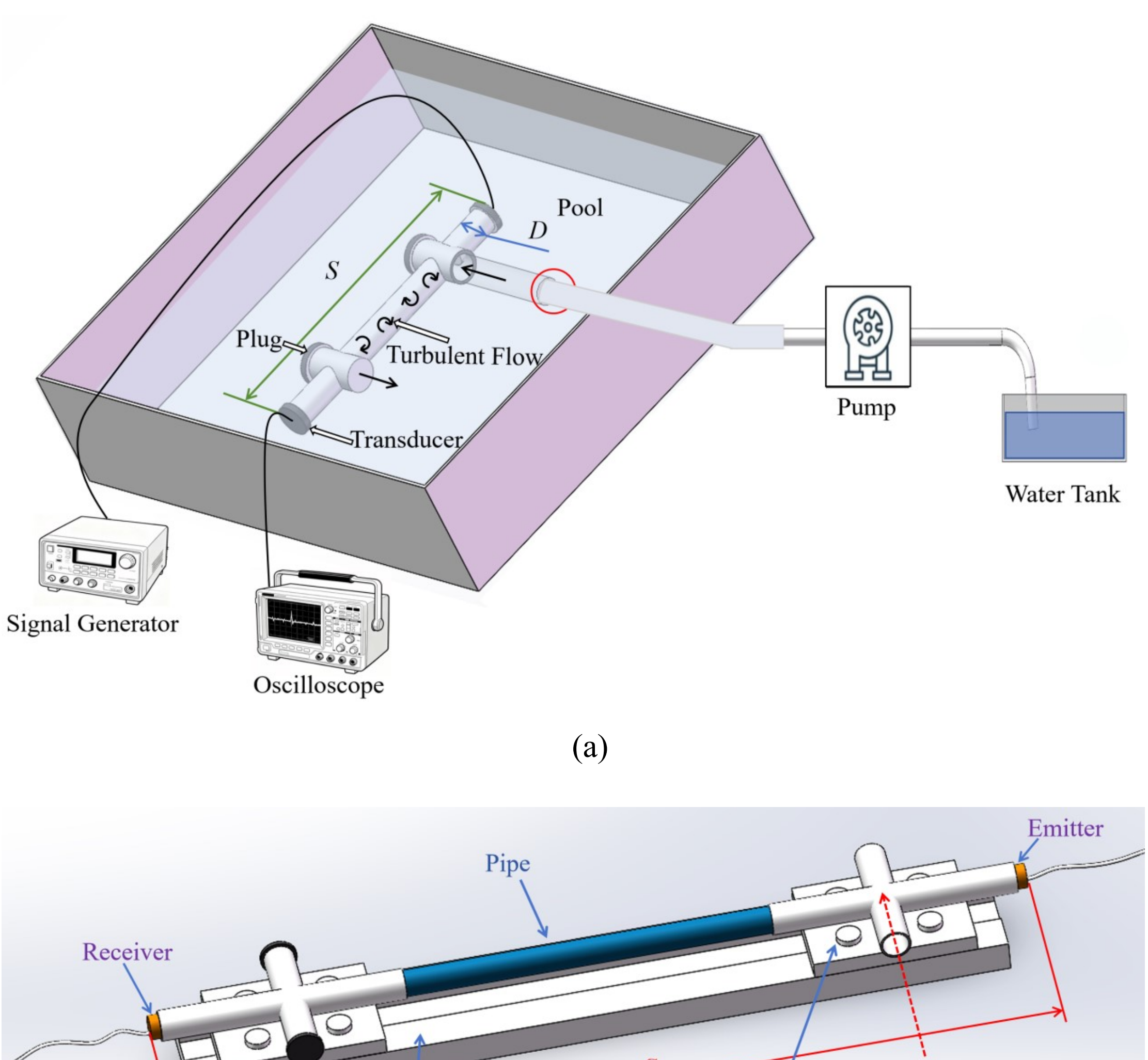

(a)

(b)

Figure 1 Schematic diagram of the experimental setup for the interaction between acoustic waves and turbulence in pipe flow, where the acoustic wave propagation direction and the mean flow direction are (a) parallel; (b) perpendicular.

The flow rates under various conditions are all approximately 7.0 L/min. Thus, for the pipe with the inner diameter $D$=1.7 cm (Figs.4~8,10~13, the fundamental frequency of transducers is $f_0$ =1, 2 MHz), the Reynolds number of the pipe flow is approximately $Re$≈9000. For the pipe with $D$=7.5 cm (Figs.7~8, $f_0$ =0.2 MHz), $Re$≈2000. For the pipe with $D$=3.66 cm (Fig.9, $f_0$ =7 kHz), $Re$≈4200. The turbulence Mach number $M_t$ has $O(M_t) \sim 10^{-5}$ (Hu & Hu 2025), suggesting that the scattering effect caused by turbulence can be neglected.

## 3. Results

### 3.1 The impact of turbulence on the phase

The Lissajous figure on the oscilloscope is used to study the effect of turbulence on the phase of acoustic waves. The principle is as follows: two signals of the same frequency are input into the oscilloscope,

$$x = A_1 \sin(\omega t + \varphi_1), \tag{1a}$$

$$y = A_2 \sin(\omega t + \varphi_2), \tag{1b}$$

where $\omega$ is the angular frequency, $t$ is time, $A_1, A_2$ are the respective amplitudes, and $\varphi_1, \varphi_2$ are the respective phases.

By setting the oscilloscope to the X-Y mode, one can observe the Lissajous figure formed by the two input signals, whose shape is determined by the phase difference $\Delta\varphi = \varphi_1 - \varphi_2$ and their respective amplitudes (Figure 2). When the phases of the two signals are the same $\Delta\varphi = 0$ , the Lissajous figure is a straight line passing through the first and third quadrants. When the phase difference is $\Delta\varphi = \pi/4$ , the Lissajous figure is an ellipse. When the phases are opposite $\Delta\varphi = \pi$ , the Lissajous figure is a straight line passing through the second and fourth quadrants.

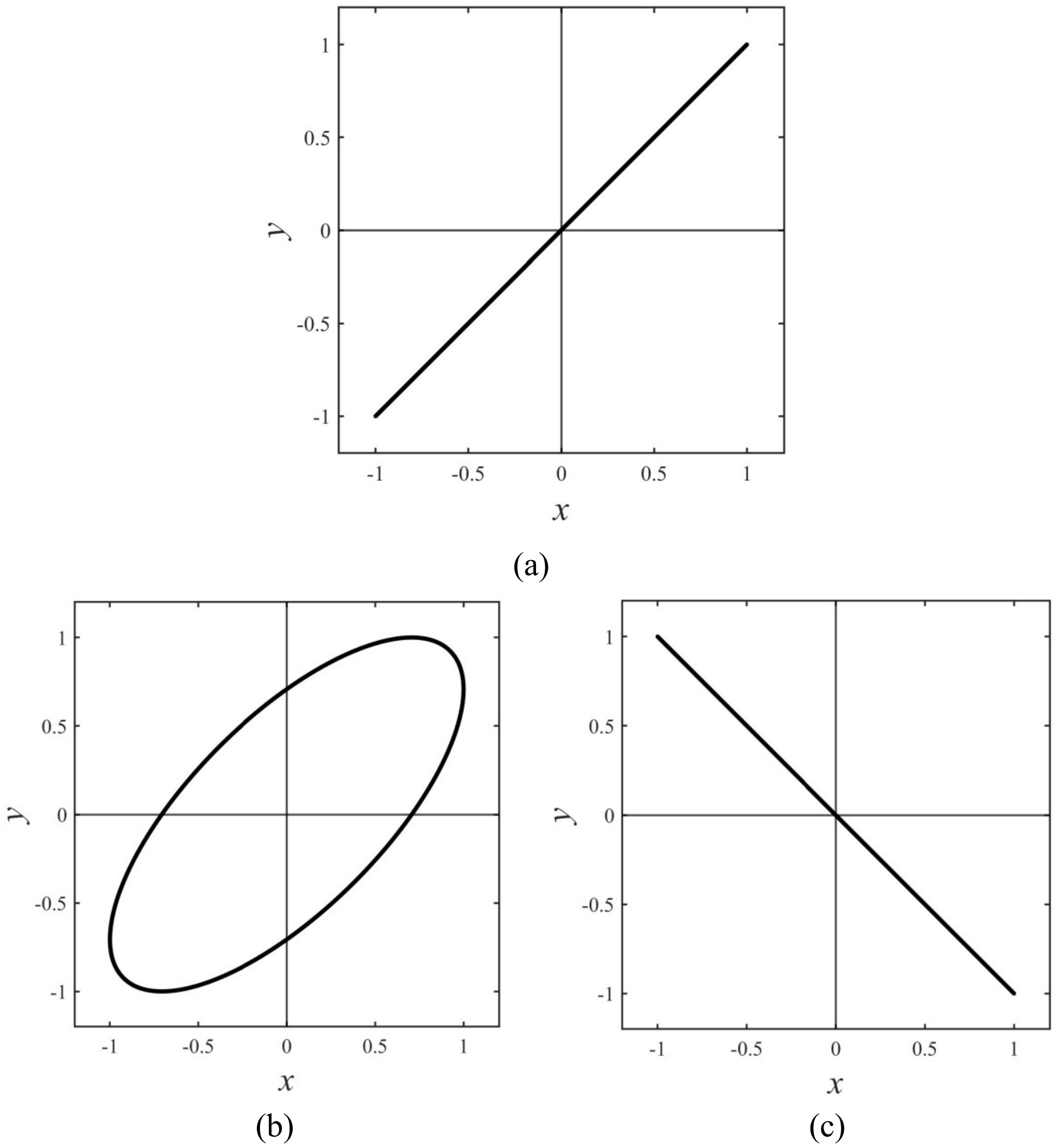

Figure 2 Lissajous figures under different phase differences: (a) $\Delta\varphi = 0$ ; (b) $\Delta\varphi = \pi/4$ ;(c) $\Delta\varphi = \pi$ . Here, $A_1 = A_2 = 1$ .

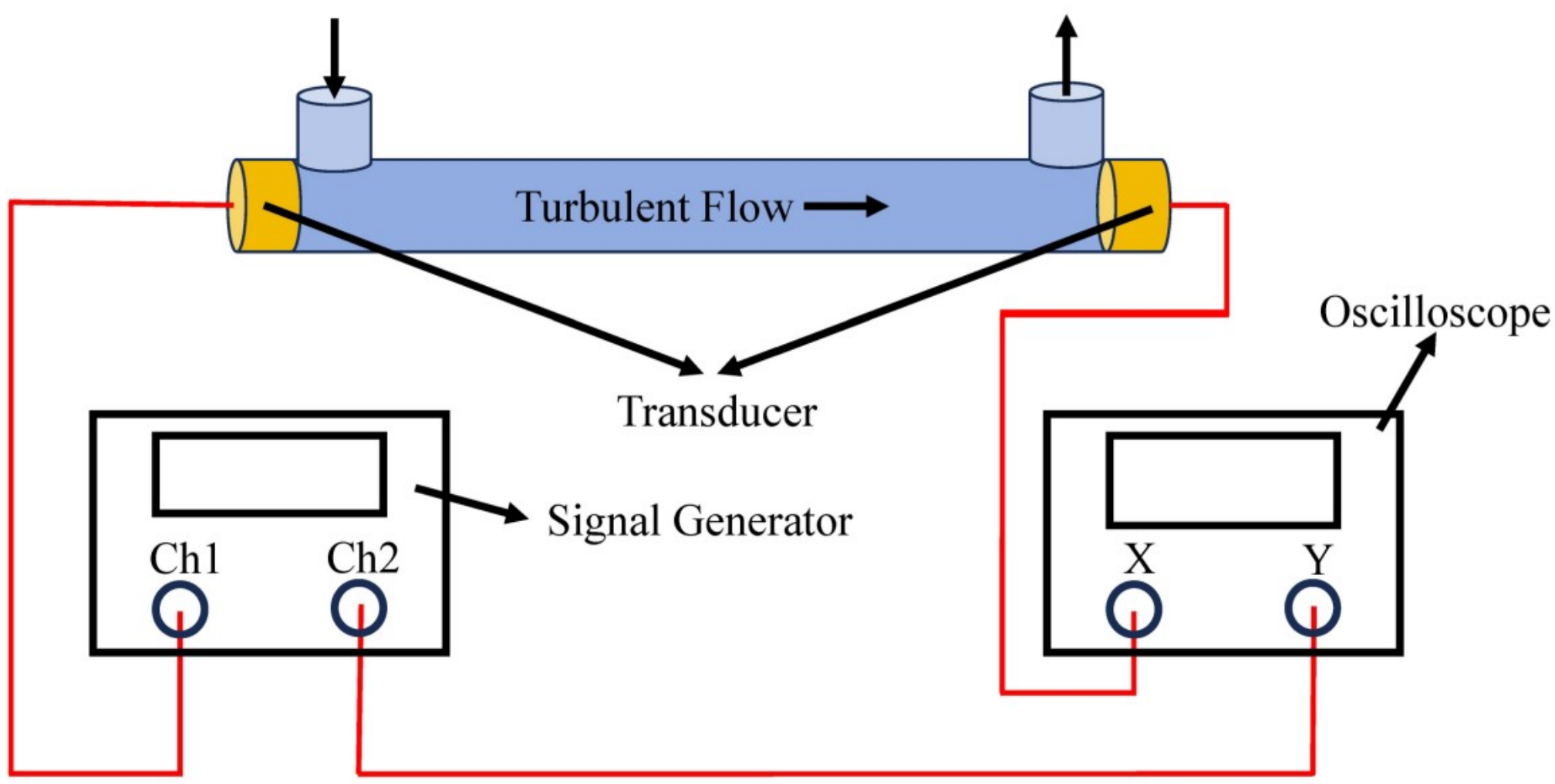

Figure 3 Schematic diagram of the experiment for measuring the phase shift of acoustic waves due to turbulence using Lissajous figures.

Based on the principles described above, the experiment in Figure 3 is conducted: First, under static water conditions, the received signal $V_r$ and another sinusoidal signal $V_n$ of the same frequency generated by the signal generator are simultaneously input into the oscilloscope to form a Lissajous figure. Then, the phase of $V_n$ is adjusted to match that of $V_r$. Subsequently, turbulence is introduced into the pipe, and the phase difference between $V_r$ and $V_n$ is observed. The experiment demonstrate that turbulence not only alters the amplitude of the acoustic wave but also changes its phase.

In Video 1, during the initial stage when the water is still, we adjust $V_r$ and $V_n$ to the same phase, resulting in a Lissajous figure that is a straight line passing through the first and third quadrants (Figure 4b). When turbulence is injected into the pipe (generated by a jet), the Lissajous figure becomes an ellipse (Figure 4c), indicating that a phase difference exists between $V_r$ and $V_n$ at this moment. After the injection stops and the mean flow in the pipe disappears, the elliptical Lissajous figure persists for some time, demonstrating that turbulent fluctuations are the cause of the phase shift. Subsequently, when the syringe is used for suction, the Lissajous figure transitions from an ellipse back to a straight line (Figure 4b), meaning the phase shift disappears. At this point, the water containing turbulent fluctuations is suctioned into the syringe, and external water enters the pipe.

Video 2 and Video 3 demonstrate that laminar flow generated by suction does not alter the phase, where the mean flow direction is parallel and perpendicular to the direction of wave propagation, respectively. This further confirms that the phase shift is indeed caused by turbulent fluctuations.

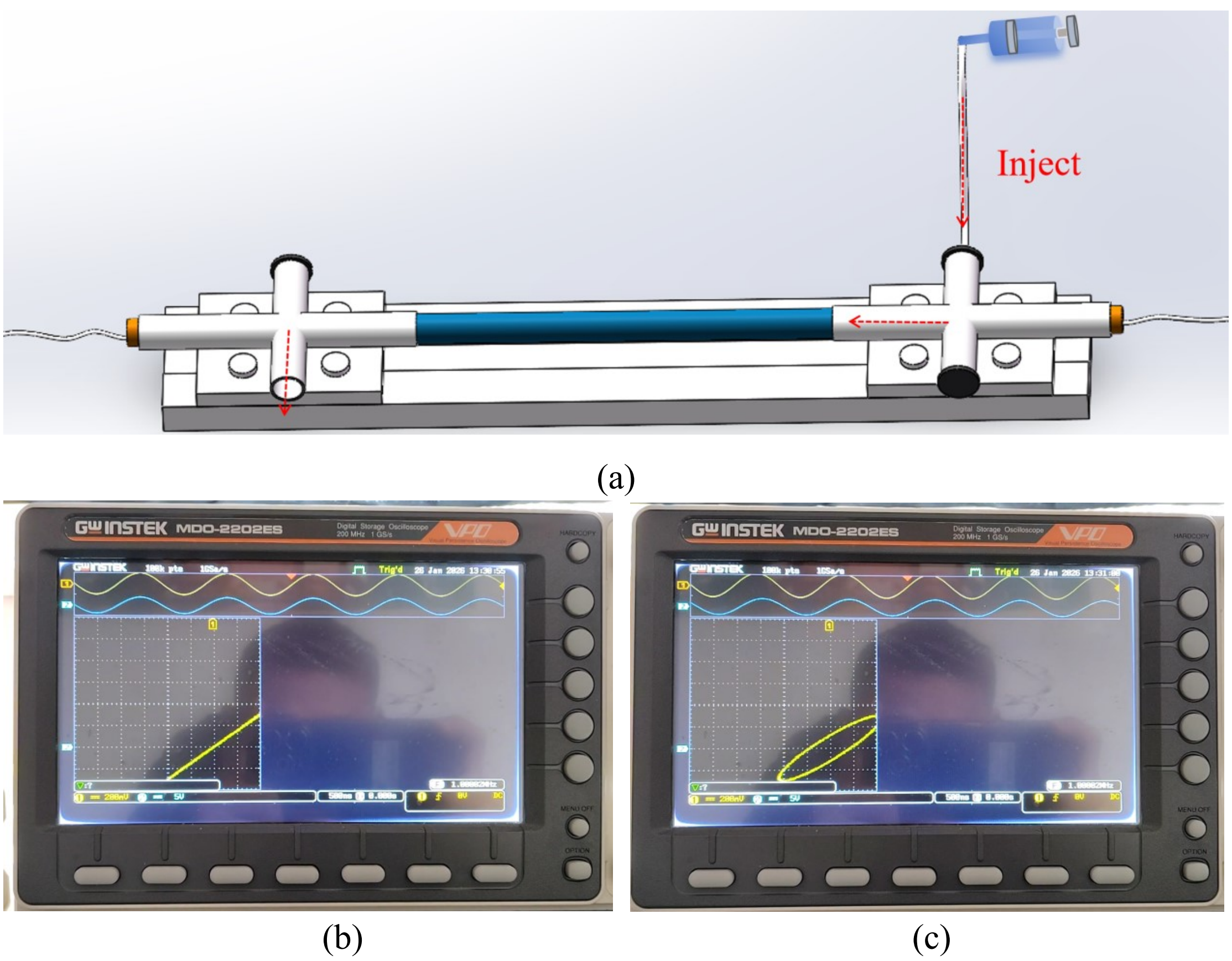

(a)

(b) (c)

Figure 4 Effect of injecting turbulence into the pipe on the phase of acoustic waves: (a) experimental setup; (b) Lissajous figure when the water is still; (c) Lissajous

figure after turbulence injection. The wave frequency is $f = 1$ MHz.

In Video 4, during the initial stage when the water is still, the Lissajous figure is adjusted to a straight line passing through the first and third quadrants. At this point, we change the voltage of the transmitted signal to modify the amplitude of $V_r$ . This only alters the slope of the line without transforming it into an ellipse. Subsequently, turbulence is injected into the pipe using a water pump (with the acoustic wave propagation direction perpendicular to the mean flow direction). At this stage, the Lissajous figure becomes an ellipse. Then, the phase of $V_n$ is adjusted to restore the Lissajous figure to a straight line passing through the first and third quadrants. The phase adjustment value of $V_n$ corresponds to the phase shift induced in $V_r$ by turbulence. After this adjustment, changing the amplitude of $V_r$ again only affected the slope of the line. This demonstrates that the phase shift is independent of the wave amplitude.

Ref.(Hu & Hu 2025) shows that when acoustic waves in a pipe are affected by turbulence, the amplitude amplification factor for the entire pipe is equal to the product of the amplification factors of the segments. Similarly, we assume that the phase shift caused by turbulence in the entire pipe is equal to the sum of the phase shifts of the segments.

To verify this hypothesis, the following experiment is conducted (Figure 5). Two water injection ports are inserted into the pipe, with flow injected separately by two water pumps, ensuring that the direction of acoustic wave are perpendicular to the mean flow direction. This setup allows for two concentrated turbulence regions in the pipe, which are approximately independent of each other.

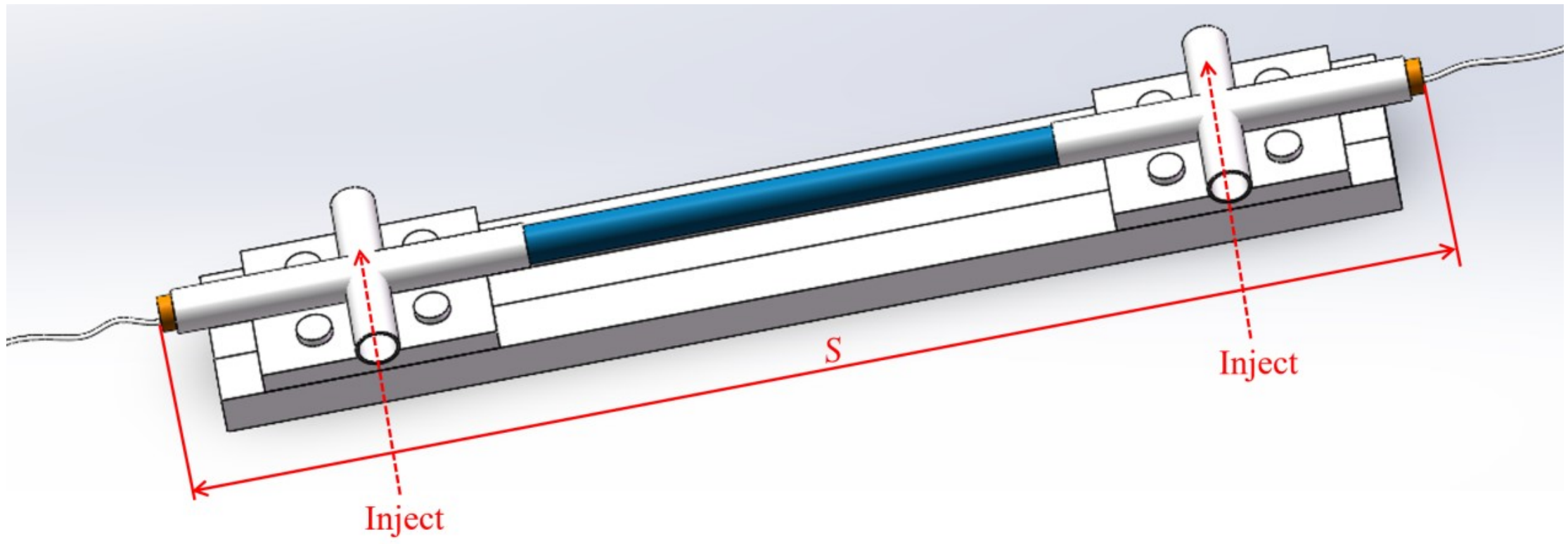

Figure 5 Experimental setup of the interaction between acoustic waves and turbulence with two water injection ports. The wave frequency is $f = 1$ MHz.

Table 1 shows the phase shift when two water pumps are activated separately and simultaneously. Among them, $\Delta\Phi_1$ ( $\Delta\Phi_2$ ) represents the phase shift when activating a single water pump to inject water into the inlet near the transmitter (receiver). $\Delta\Phi_{12}$ represents the phase shift when both water pumps are activated simultaneously. By comparing the relative errors between the theoretical value $\Delta\Phi_1 + \Delta\Phi_2$ and the actual value $\Delta\Phi_{12}$ , one can find that they are very close, thereby confirming the earlier conjecture.

In addition, we also observe the phase shift caused by jet turbulence in an open environment, as shown in Figure 6, where the direction of the acoustic wave is perpendicular to the direction of the jet. The results (video 5) are similar to those in Figure 4 . When the jet occurs, the Lissajous figure changes from a straight line to an ellipse, indicating that the jet turbulence alters the phase.

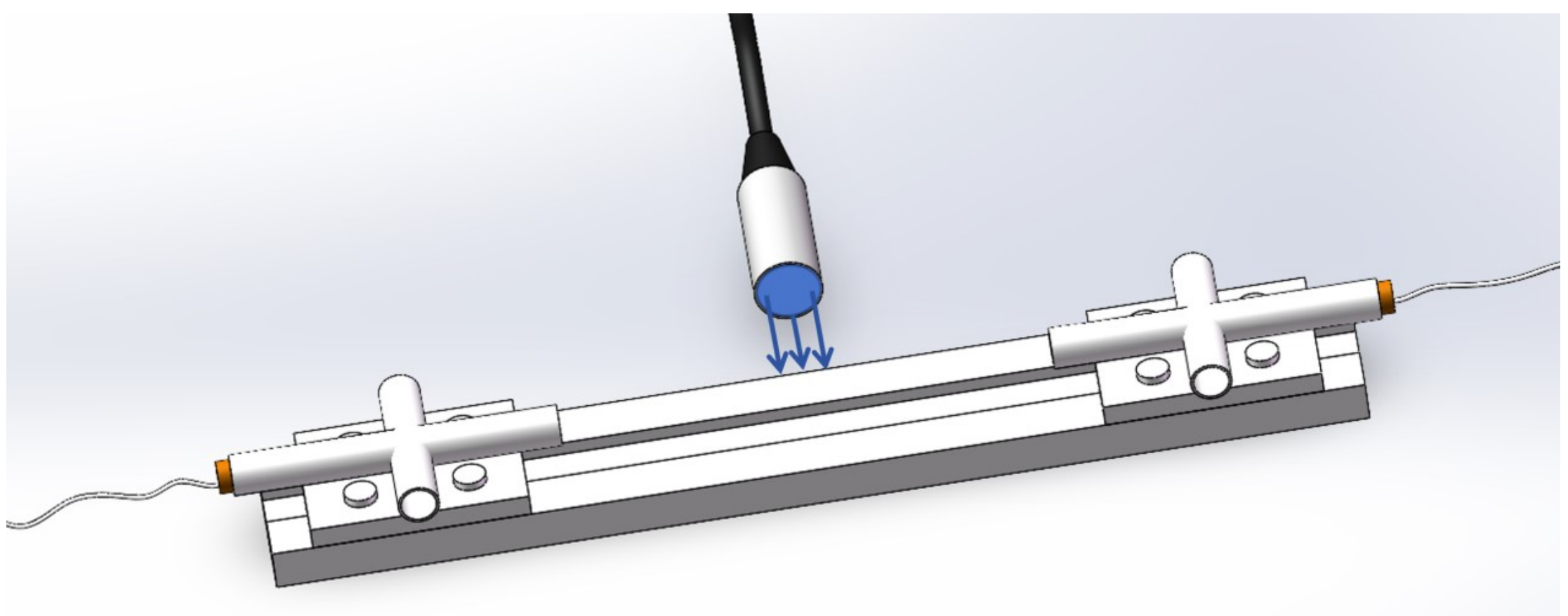

Figure 6 Experimental setup of the interaction between acoustic waves and jet turbulence in an open environment. The wave frequency is $f$ = 1 MHz.

**Table 1** The phase shifts of acoustic waves in a pipe with two water injection ports under various conditions. Among them, the voltage of the transmitted signal output to the transducer is 10 V, and the voltage of the signal connected to the oscilloscope is 15 V.

| No. | Temperature (°C) | Emitter | Receiver | | | | |
|---|---|---|---|---|---|---|---|
| | | Frequency (MHz) | $\Delta\Phi_1$ (°) | $\Delta\Phi_2$ (°) | $\Delta\Phi_{12}$ (°) | $\Delta\Phi_1+\Delta\Phi_2$ (°) | $\left\|\Delta\Phi_1+\Delta\Phi_2-\Delta\Phi_{12}\right\|/\Delta\Phi_{12}$ |
| 1 | 10.3 | 1.00 | 23±1 | 30.0±1 | 54±1 | 53±2 | 1.8% |
| 2 | 8.9 | 1.05 | 22±1 | 35.0±1 | 58±1 | 57±2 | 1.7% |
| 3 | 10.7 | 1.10 | 34±1 | 52.0±1 | 86±1 | 86±2 | 0% |
| 4 | 8.9 | 1.15 | 17±1 | 25.0±1 | 44±1 | 42±2 | 4.5% |

Therefore, turbulence simultaneously alters both the amplitude and phase of acoustic waves. The total phase shift across the entire pipeline equals the sum of the phase shifts of the segments of the pipeline, while the total amplification factor equals the product of the amplification factors of the segments. This property reminds us of the relationship between amplitude and phase changes induced by the complex refractive index in laser gain media.

In terms of the complex refractive index $n = n_R + \mathrm{i} n_I$ , a incident plane wave experiences amplification (or attenuation) and a phase shift determined by $n_I$ and $n_R$ , respectively (Milonni 2010, p759):

$$\exp(-\mathrm{i}kz) = \exp\left(-\mathrm{i}\frac{\omega(n_R + \mathrm{i}n_I)}{C_0}z\right) = \exp\left(\frac{\omega n_I z}{C_0}\right)\exp\left(-\mathrm{i}\frac{\omega n_R z}{C_0}\right), \tag{2}$$

where $C_0$ is the speed of light in vacuum. Here, the propagation direction of the incident wave is taken as the positive $z$-axis. The amplification factor $A$ and the phase shift $\Delta\Phi$ relative to vacuum are

$$A = \exp\left(\frac{\omega n_I \Delta z}{C_0}\right), \quad \Delta\Phi = \frac{\omega(n_R - 1)\Delta z}{C_0}, \tag{3}$$

where $\Delta z$ is the length along $z$-axis. Then, the total amplification factor equals the product of the amplification factors of the segments,

$$\exp\left[\frac{\omega n_I(\Delta z_1+\Delta z_2)}{C_0}\right]=\exp\left(\frac{\omega n_I\Delta z_1}{C_0}\right)\cdot\exp\left(\frac{\omega n_I\Delta z_2}{C_0}\right), \tag{4}$$

and the total phase shift of the wave in the entire pipeline equals the sum of the phase shifts of the segments,

$$\frac{\omega(n_R-1)(\Delta z_1+\Delta z_2)}{C_0}=\frac{\omega(n_R-1)\Delta z_1}{C_0}+\frac{\omega(n_R-1)\Delta z_2}{C_0}, \tag{5}$$

where $\Delta z_1, \Delta z_2$ are the axial lengths of the two segments, respectively. These properties are completely consistent with the changes in acoustic waves within turbulence.

The incident wave undergoes changes in amplitude and phase before reflecting at the reflection surface. Therefore, for the reflected wave, the same amplification factor and phase shift are generated at the origin. Consequently, the amplification factor and phase shift of the superposition of the incident and reflected waves are also consistent with the above derivation.

In the experiment shown in Figure 3, increasing the phase of the X-channel signal causes the Lissajous figure to change from a straight line passing through the first and third quadrants into an ellipse, while increasing the phase of the Y-channel signal causes the ellipse to return to a straight line. This indicates that if we analogize the effect of turbulence on acoustic waves to the refraction of a medium, its refractive index would lead to an increase in phase, therefore $n_R < 1$.

### 3.2 The Variations of Amplification factor and Phase shift with Frequency

In the following, $V_1$ is the transmitted signal voltage, while $V_2$, and $V_3$ represent the received signal voltage under static water and turbulent flow, respectively. The relative changes is $V_a = (V_3 / V_2 - 1)$. and the amplification factor is defined as $A=V_3/V_2$. It is found that the amplification factor caused by turbulence is closely related to frequency. Thus, using the experimental setup in Figure 1(b), the amplification factors at different frequencies are measured for the case where the acoustic wave is perpendicular to the mean flow.

Figure 7 shows that the amplification factor at various frequencies exhibits periodic variations as the frequency increases. In Figure 7(a), the frequencies are relatively close to each other, resulting in approximately equal periods of variation in the amplification factor. In contrast, in Figure 7(b), the periods near 0.2 MHz and 2 MHz show noticeable differences. This indicates that the period of variation in the amplification factor is related to the frequency.

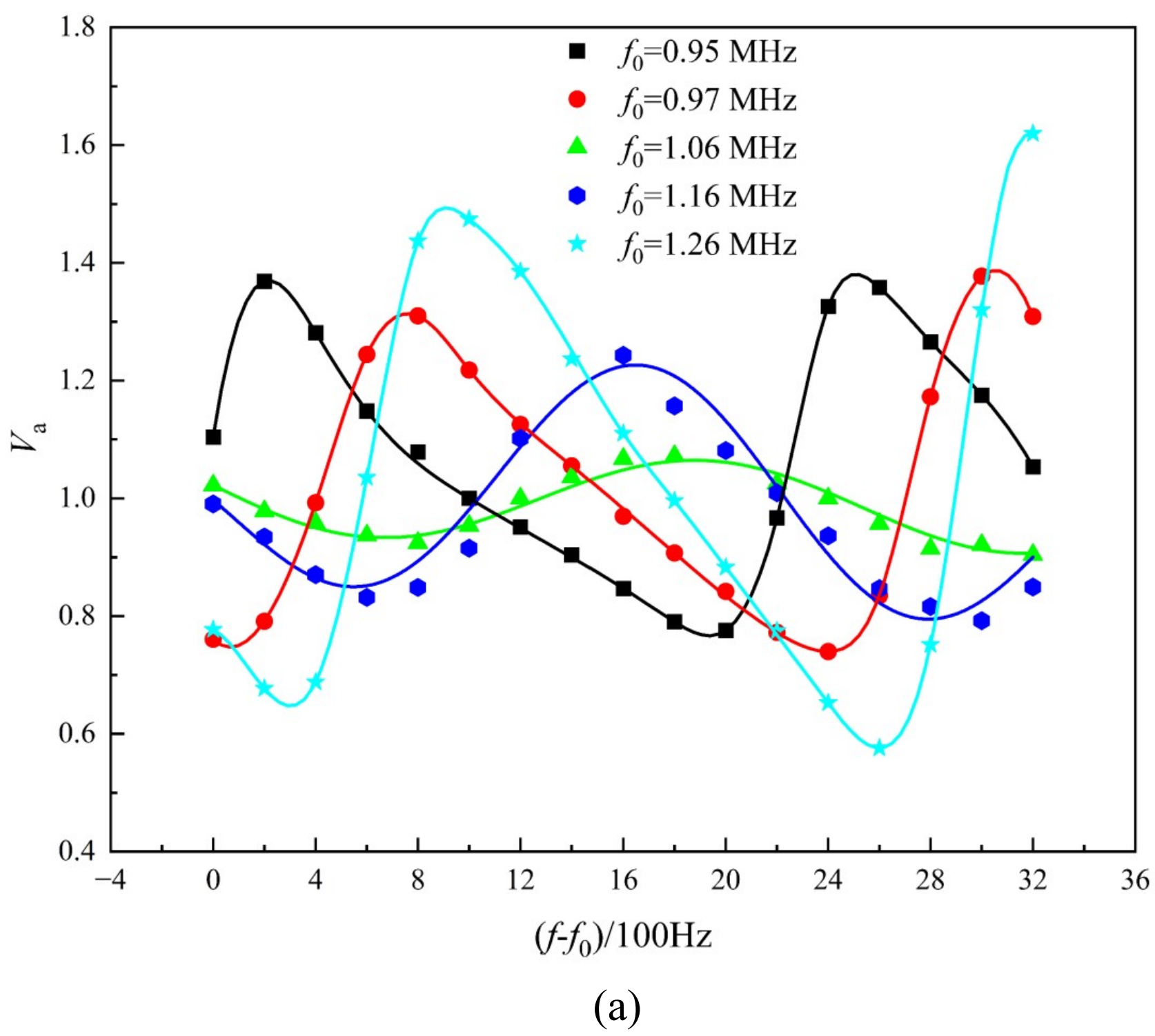

(a)

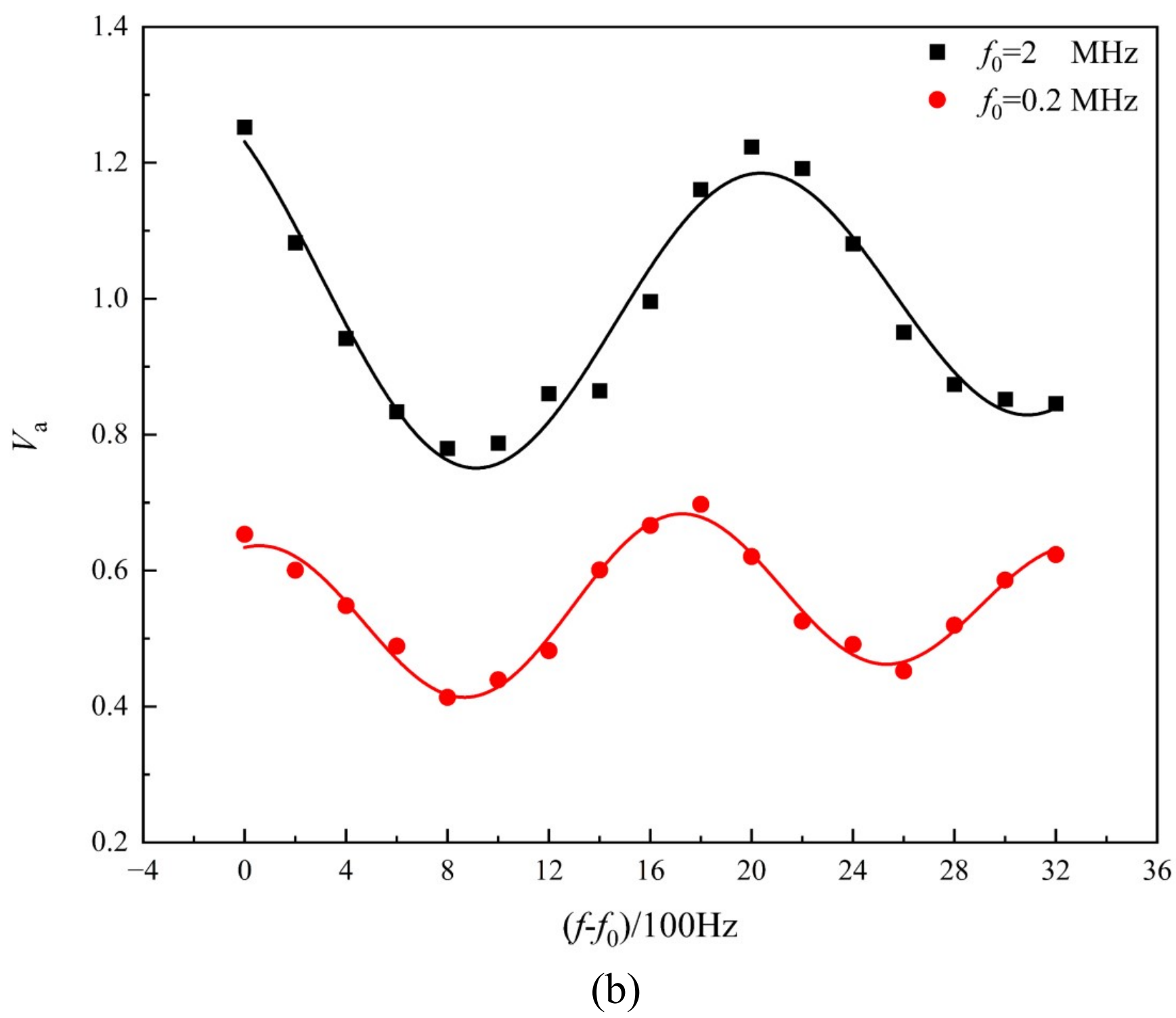

(b)

Figure 7 Amplification factor at different frequencies generated by turbulence inside the pipeline, where the acoustic wave direction is perpendicular to the mean turbulent flow. The water temperature is (11.8±0.2)°C. The standard uncertainty of $V_a$ is approximately 1%.

This periodic behavior reminds us of similar phenomena observed in stimulated emission. The total amplification factor of a laser can be approximately expressed as:

$$M(\nu)=g(\nu)\cdot T(\nu), \tag{6}$$

where $g(\nu)$ is the gain coefficient of the gain medium (Milonni 2010, p115), typically a unimodal function (such as a Lorentzian or Gaussian lineshape), and $T(\nu)$ is the transmission function of the resonant cavity(Milonni 2010, p225), which can be periodic. For example, in a Fabry-Pérot etalon, as light undergoes multiple reflections between the two mirrors of the cavity, only optical frequencies that satisfy the resonance condition can form stable oscillations (Milonni 2010, p201):

$$\nu_m = m\frac{c}{2nL}, \tag{7}$$

where $m$ is an integer, $L$ is the cavity length, and $n$ is the refractive index. The total amplification factor $M(\nu)$ exhibits peaks at these discrete frequencies, with the spacing between adjacent peaks given by:

$$\Delta\nu = \frac{c}{2nL}. \tag{8}$$

This may result in periodic fluctuations, but the underlying cause is the filtering effect of the resonant cavity, rather than the physics of stimulated emission itself.

In Figure 7(a), the spacing $\Delta\nu$ between adjacent peaks of the amplification factor at various frequencies is approximately 2300 Hz. In Figure 7(b), for $f_0$= 0.2 MHz and 2 MHz, $\Delta\nu$ are about 1800 Hz and 2300 Hz, respectively.

Given the speed of sound $C \approx 1454\text{m/s}$ in water at 11.8°C and a cavity length of $L$ =32cm, substituting these values into Equation (8) yields results $\Delta\nu \approx 2272\text{Hz}$, that are relatively consistent with the experimental values in the 1 MHz to 2 MHz frequency range. However, for the 0.2 MHz case, the periodicity of the amplification factor may be somewhat influenced by the gain coefficient of the medium.

Figure 8 shows that the phase shift also exhibits periodic oscillations as the frequency increases, but its oscillation period is clearly different from that of the amplitude oscillation. For $f_0$= 0.2 MHz, 1 MHz, 1.3 MHz, and 2 MHz, the spacings $\Delta\nu$ between adjacent peaks of the phase change are approximately 1400 Hz, 3400 Hz, 2400 Hz, and 2800 Hz, respectively.

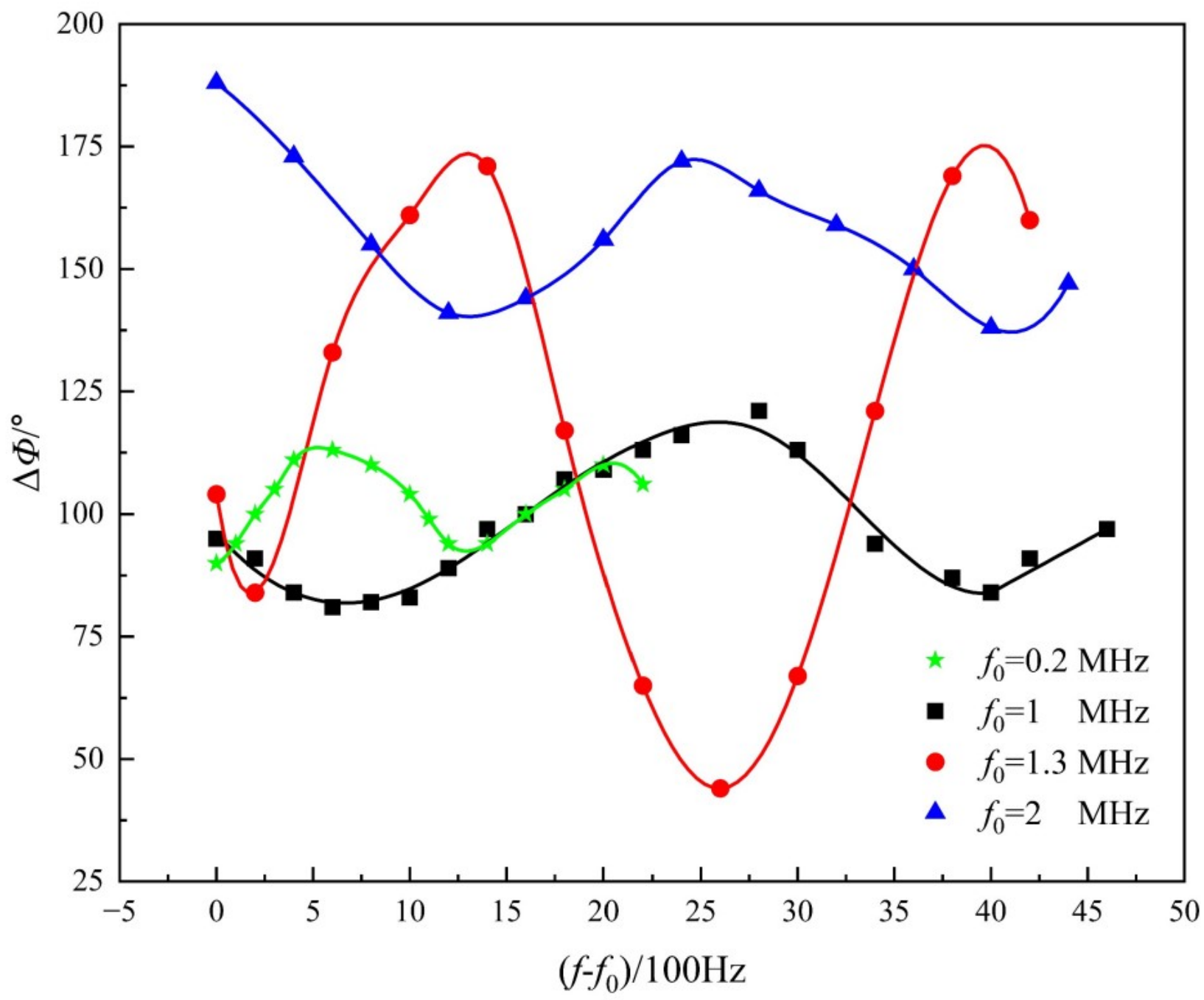

Figure 8  The phase shift at different frequencies generated by turbulence inside a pipe, where the direction of the acoustic wave is the same as the mean flow direction. The water temperature is (9.2 ± 0.7)°C. The standard uncertainty of $\Delta\Phi$ is approximately 1°.

### 3.3 Temporal evolution of acoustic waves

When the pump is shut off after the pipe flow has stabilized, the turbulence gradually decays thereafter. The received signal undergoes a transient variation process and eventually converges to the value observed under static conditions. This process reflects the influence of decaying turbulent fluctuations on the acoustic wave.

Table 2 shows six types of amplitude variations of the received acoustic signal over time in pipe flow. Under the condition of the transmitted signal amplitude $V_1$, the received signal voltages $V_r$ include $V_2$, $V_3$, $V_4$. These are amplitudes of the received signals under three conditions: when the water is stationary, the steady-state value after the pump is started (Stable Turbulence), and the extreme value achievable after the pump is turned off (Decaying Turbulence).

It can be seen that under different acoustic frequencies, the magnitudes of $V_2$, $V_3$, $V_4$ can exhibit various comparative relationships. Since different turbulence scales decay at different rates during the turbulence attenuation process, the received acoustic signal reflects the combined effect of all turbulence scales on the acoustic wave. This indicates that turbulence fluctuations of different scales have distinct amplification factors on the acoustic wave.

However, during the turbulence decay process, the phase shift monotonically decreases to zero over time without non-monotonic variations.

### 3.4 Low-frequency Acoustic Waves

We conduct experiments on the interaction between turbulence and low-frequency acoustic waves (100Hz ~ 7kHz) under two conditions: with the pipe placed in air (Figure 9) and submerged in water (Figure 1a). For the former, both amplitude and phase are measured, while for the latter, the signal waveform in the time domain is not stable enough to allow precise phase measurements. Therefore, only the amplitude in the frequency domain is measured.

Table 3 shows that when the pipe is placed in air and contains no water, the receiver can still detect a signal of a certain amplitude, but this amplitude is significantly smaller than when the pipe is filled with static water. When turbulence is introduced into the pipe, the change in acoustic wave amplitude is within the precision range of the oscilloscope, essentially indicating no observable change. Observations from the

**Table 2** The types of temporal evolution of the received signal in pipe turbulence, where the acoustic wave propagates in the same direction as the mean flow.

| No. | Type | Temperature (°C) | Emitter | | Receiver | | |
|---|---|---|---|---|---|---|---|
| | | | Frequency (MHz) | Voltage $V_1$(V) | Static $V_2$(mV) | Stable Turbulence $V_3$(mV) | Decaying Turbulence $V_4$(mV) |
| 1 | | 13.5 | 0.9400 | 5 | 13.20±0.04 | 32.83±0.47 | — |

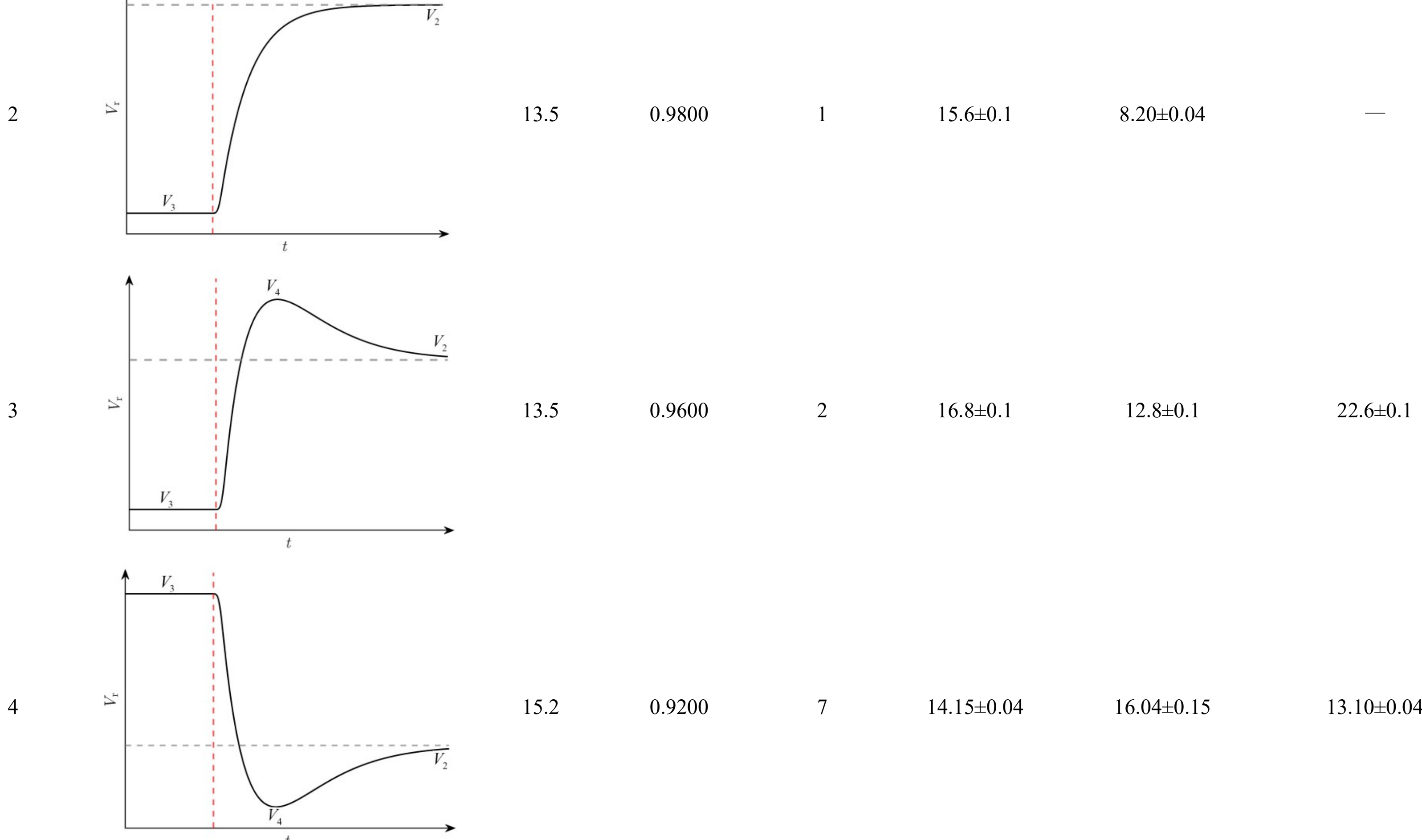

| | | | | | | | |
|---|---|---|---|---|---|---|---|
| 2 | | 13.5 | 0.9800 | 1 | 15.6±0.1 | 8.20±0.04 | — |
| 3 | | 13.5 | 0.9600 | 2 | 16.8±0.1 | 12.8±0.1 | 22.6±0.1 |
| 4 | | 15.2 | 0.9200 | 7 | 14.15±0.04 | 16.04±0.15 | 13.10±0.04 |

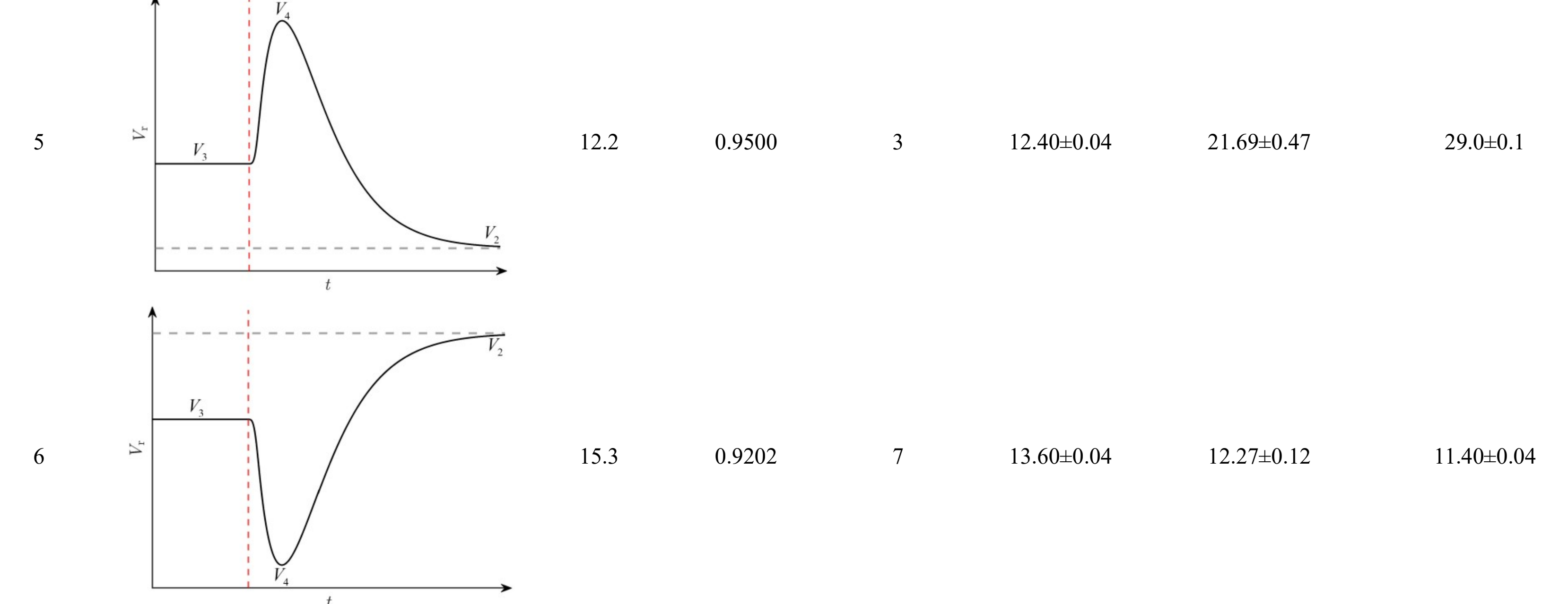

| | | | | | | | |
|---|---|---|---|---|---|---|---|
| 5 | | 12.2 | 0.9500 | 3 | 12.40±0.04 | 21.69±0.47 | 29.0±0.1 |
| 6 | | 15.3 | 0.9202 | 7 | 13.60±0.04 | 12.27±0.12 | 11.40±0.04 |

Lissajous figures further indicate that turbulence does not alter the phase.

For the case where the pipe is submerged in water, the influence of turbulence on the wave amplitude is also negligible—far smaller than that observed under the same conditions (pipe turbulence driven by a water pump, with the acoustic wave propagating in the same direction as the mean flow) for acoustic waves at frequency $f$ ≈1 MHz. Lissajous figures show that, due to the presence of noise, the waveform is highly unstable at this point, making it difficult to observe any meaningful phase shift.

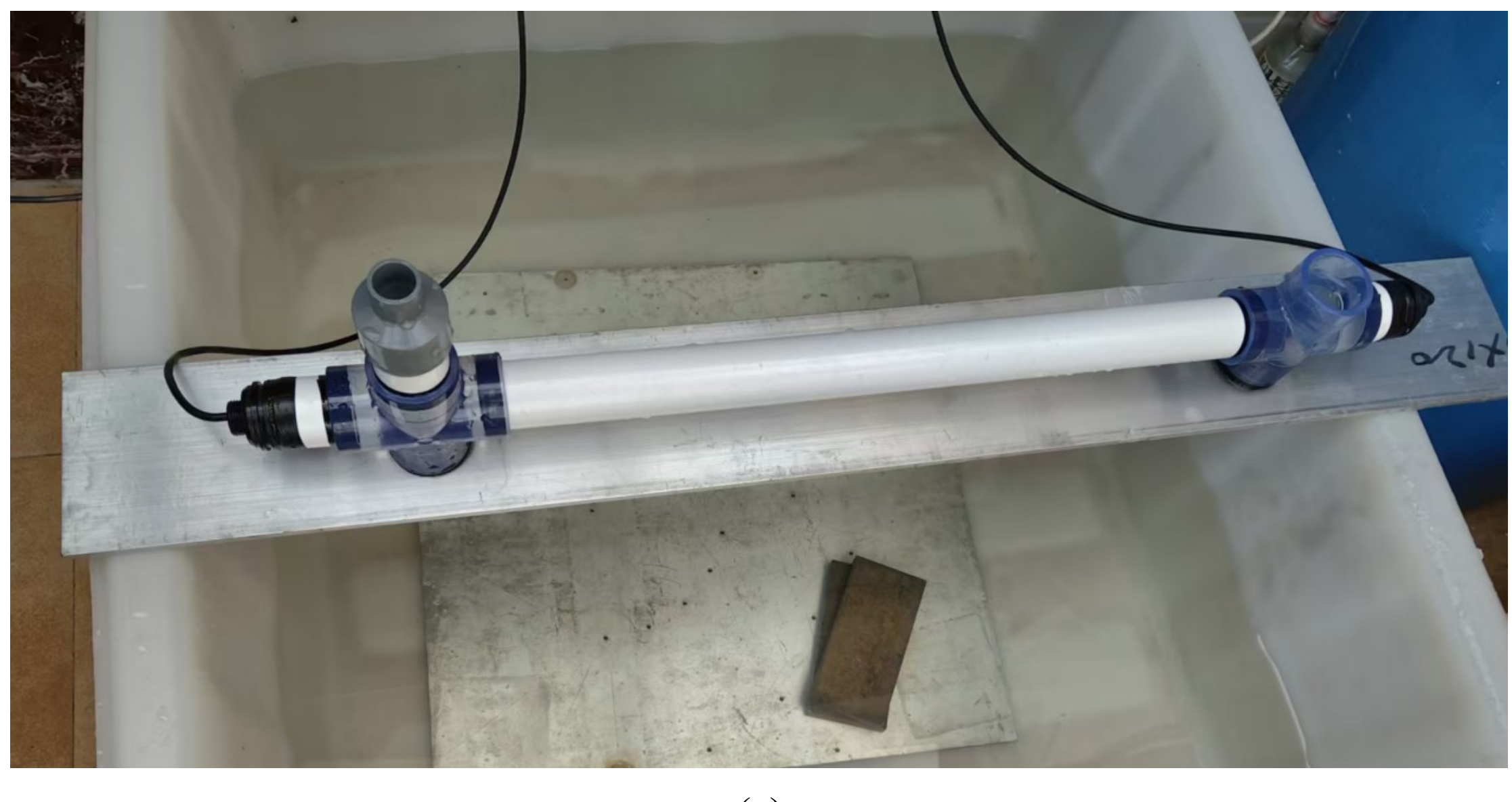

(a)

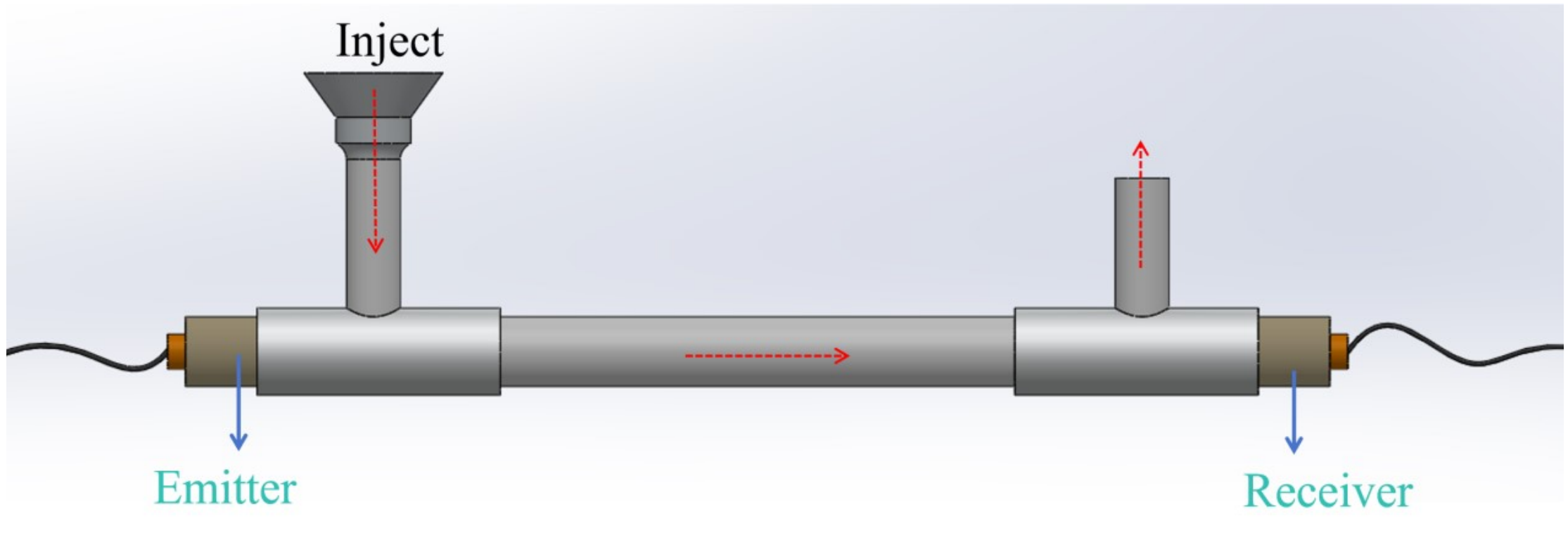

(b)

Figure 9 Experimental setup for the interaction between turbulence and low-frequency acoustic waves with the pipe placed in air: (a) actual photograph; (b) schematic diagram. The direction of acoustic wave propagation is the same as the mean flow direction.

**Table 3** Voltage signal values for the interaction between turbulence and low-frequency acoustic waves within the pipe: (a) pipe placed in air; (b) pipe submerged in water. The uncertainty in the numerical values correspond to the precision of the oscilloscope.

(a)

| No. | Temperature (°C) | Emitter | | Receiver | | |
|---|---|---|---|---|---|---|
| | | Frequency (kHz) | Voltage $V_1$(V) | No water $V_0$(mV) | Static $V_2$(mV) | Turbulence $V_3$(mV) |
| 1 | 10.3 | 7 | 10 | 1.28±0.04 | 28.1±0.1 | 28.2±0.1 |
| 2 | 7.3 | 3 | 5 | 2.40±0.04 | 16.4±0.1 | 16.4±0.1 |
| 3 | 11.5 | 1 | 10 | 1.12±0.04 | 27.0±0.1 | 27.0±0.1 |
| 4 | 7.4 | 0.5 | 10 | 3.84±0.04 | 19.4±0.1 | 19.4±0.1 |
| 5 | 7.5 | 0.2 | 15 | 2.64±0.04 | 21.5±0.1 | 21.4±0.1 |
| 6 | 7.6 | 0.1 | 20 | 2.40±0.04 | 17.8±0.1 | 17.8±0.1 |

(b)

| No. | Temperature | Emitter | Receiver |
|---|---|---|---|

|  | (°C) | Frequency | Voltage | Static | Turbulence |
|---|---|---|---|---|---|
|  |  | (kHz) | $V_1$(V) | $V_2$(mV) | $V_3$(mV) |
| 1 | 11.5 | 7 | 10 | 40.2±0.2 | 40.0±0.2 |
| 2 | 11.4 | 3 | 10 | 37.4±0.2 | 37.2±0.2 |
| 3 | 11.5 | 1 | 10 | 30.8±0.2 | 30.6±0.2 |
| 4 | 11.5 | 0.5 | 10 | 22.2±0.1 | 22.2±0.1 |
| 5 | 11.7 | 0.2 | 15 | 16.0±0.1 | 16.0±0.1 |
| 6 | 11.7 | 0.1 | 20 | 10.1±0.1 | 10.1±0.1 |

The above results indicate that for acoustic waves with frequencies below 7 kHz, their amplitude and phase are not altered by turbulence.

### 3.5 High-frequency Acoustic Waves

To investigate the influence of turbulence on high-frequency acoustic waves, a square wave signal is used in the signal generator, while the receiver's spectrum is displayed in logarithmic form (Figure 10). The voltage of the transmitted signal is 20V, the duty cycle of the square wave is 50%, and the frequency is $f_0$=1.16MHz. Under the fundamental frequency $f_0$, spectral lines are observed at frequencies $f=mf_0$ in the received signal, where $m$ is a natural number. Figure 10 shows that in static water, the spectrum of the received signal still exhibits spectral lines around 25 MHz, significantly above the noise level.

In Video 6, the fundamental frequency of the square wave signal is $f_0$ =1.06 MHz. When turbulence is introduced into the pipe, the effect of turbulence on the amplitude of acoustic waves is significant for frequencies below 6 MHz. However, for waves with frequencies above 10 MHz (Video 7), the amplitude variation is negligible. Additionally, we test square wave signals with various frequencies, including $f_0$ =1.13,1.16,1.21 MHz, and the results are similar. This indicates that turbulence has little effect on high-frequency acoustic waves.

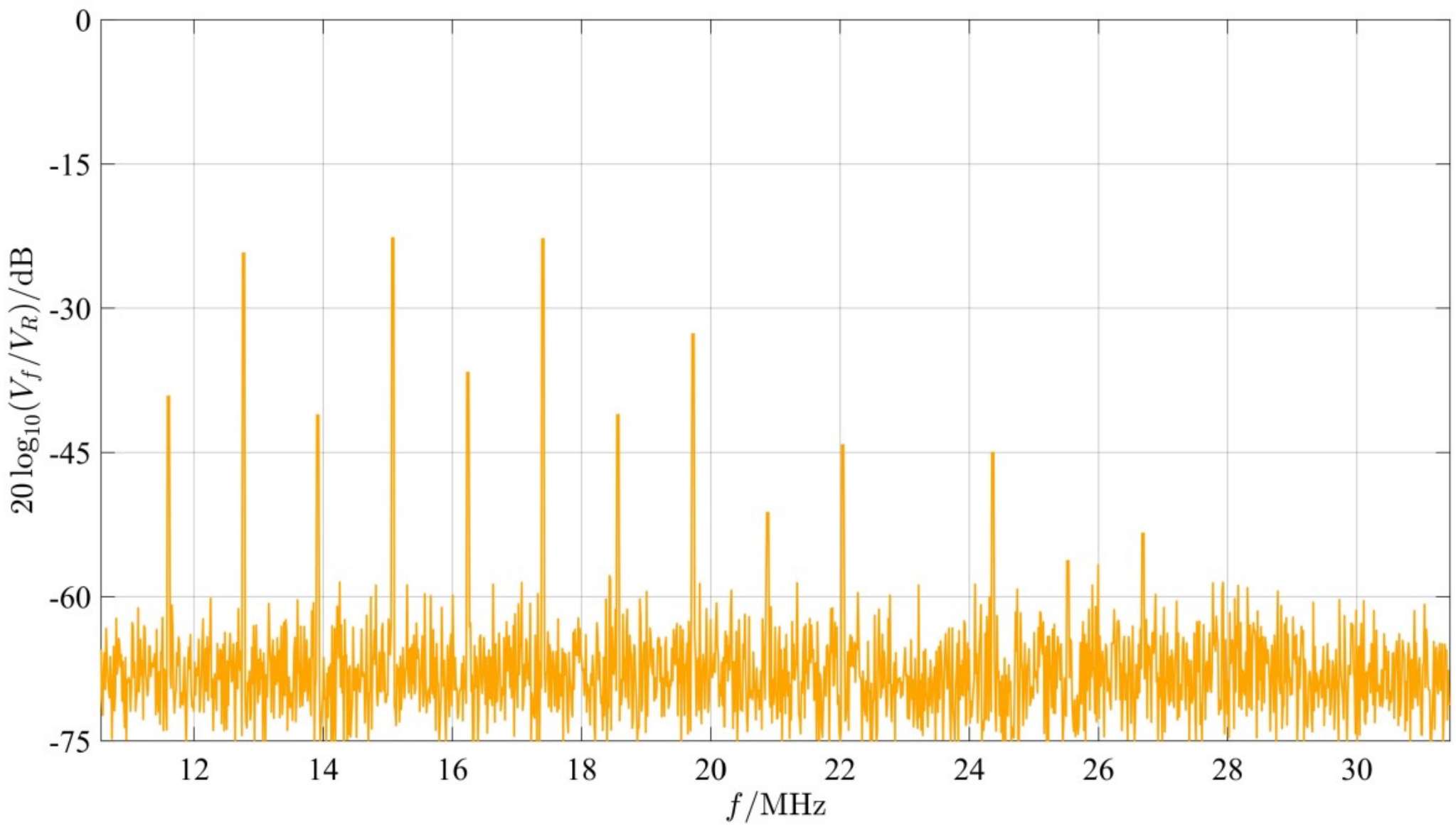

Figure 10 The spectrum of a square wave signal.The fundamental frequency is $f_0$ =1.16MHz. Here, $V_f$ is the amplitude of the received signal at the frequency $f$ and $V_R$ is the reference amplitude.

Based on the experiments in Sections 3.4 & 3.5, one can find that the effect of turbulence on acoustic waves exists only within a certain frequency range. Acoustic waves with frequencies below or above specific thresholds are almost unaffected by turbulence. Within this frequency range, the amplification factor and phase shift induced by turbulence vary continuously with the frequency. This phenomenon is quite analogous to the optical properties of semiconductors.

The absorption coefficient of semiconductors for light is given by $\alpha_0 \propto \sqrt{\hbar\omega - E_g}$, where $\omega$ and $\hbar\omega$ are the photon frequency and energy, respectively, $\hbar$ is Planck's constant, and $E_g$ is the bandgap energy (Jena 2022, p731). When $\hbar\omega > E_g$, the absorption coefficient exists and varies continuously with the incident photon energy, thereby defining a continuous absorption spectrum.

Philipp & Ehrenreich (1963) measured the spectral characteristics of various semiconductor materials. The results show that the imaginary parts of the dielectric constant (reflecting the degree to which the material absorbs and dissipates light) of multiple semiconductors tend to zero at both low and high frequencies, while exhibiting peaks between the two. This indicates that semiconductors possess a continuous, finite spectral response range.

In addition, semiconductors not only exhibit phenomena of stimulated absorption and emission, but they themselves constitute one of the most important classes of laser gain media, widely used in laser diodes and optical amplifiers. Combining the analogy between the effects of turbulence on acoustic wave and the complex refractive index in laser gain media in Section 3.1, the continuous, finite spectral response range observed here further demonstrates the high degree of similarity between turbulence and laser gain media.

### 3.6 Superposition of standing waves and traveling waves

In the experiment, we place two transducers facing each other to transmit and receive acoustic waves, respectively. The incident wave emitted by the transmitter can be reflected by the receiver, so the acoustic wave in the flow field is the superposition of the incident wave and the reflected wave. When the amplitudes of these two waves are equal, the resulting superimposed wave is a standing wave. However, part of the energy of the incident wave can be absorbed by the receiver, so the actual acoustic wave in the flow field can be regarded as a superposition of a traveling wave and a standing wave.

Consider a plane wave propagating along the axis of a pipe with a uniform cross-sectional area, and ignore the dissipation caused by viscosity. Let the forms of the waves be

$$p_i = p_1 \exp\left[\mathrm{i}\left(\omega t - kz\right)\right], \tag{9a}$$

$$p_r = p_2 \exp\left[\mathrm{i}\left(\omega t + kz\right)\right], \tag{9b}$$

where $p_i, p_r$ represent the sound pressures of the incident wave and the reflected wave, respectively; $k$ is the wave number, $\omega$ is the angular frequency; $z$ is the axial coordinate, with the reflecting surface taken as the origin; $p_1, p_2$ are their respective complex amplitudes. The ratio of the two is the reflection coefficient for the pressure (Skudrzyk 1971, p296),

$$r_p = p_2 / p_1 = \left|r_p\right| \exp\left(\mathrm{i}\,\theta\right). \tag{10}$$

Superimposing the incident wave and the reflected wave yields the total sound pressure within the pipe,

$$p_a = p_i + p_r, \tag{11a}$$

$$\left|p_a\right| = \left|p_1\right| \left|\sqrt{1 + \left|r_p\right|^2 + 2\left|r_p\right| \cos\left(2kz + \theta\right)}\right|. \tag{11b}$$

From equation (11b), it can be observed that when $2kz + \theta = 2m\pi, m \in Z$ , the magnitude of the total sound pressure reaches a maximum; when

$2kz+\theta=(2m+1)\pi, m\in Z$, it reaches a minimum. The distance between two adjacent maxima (or minima) is

$$\Delta z = 2\pi/(2k) = \lambda/2. \tag{12}$$

Here $\lambda$ is the wavelength. The ratio of the maximum value to the minimum value is the standing wave ratio $G$ (Skudrzyk 1971, p306):

$$G=\frac{1+|r_p|}{1-|r_p|}. \tag{13}$$

In order to study the standing wave ratio, one can measure the variation of the received signal amplitude $V_2$ with the transducer spacing $S$ in both an open environment and within the pipe for static water.

In Figure 11, we fix the two transducers on rail sliders, and move the sliders allowed for continuous adjustment of the transducer spacing. The water temperature is approximately 10°C, at which the speed of sound is $C\approx1447.2$m/s. The acoustic frequency used is $f$=1MHz, and according to the wavelength relation $\lambda=C/f$, the wavelength is $\lambda\approx1.45\text{mm}$. Therefore, a slider with a precision of 0.1mm (Figure 14) is used and the total change in slider distance is 1.5mm.

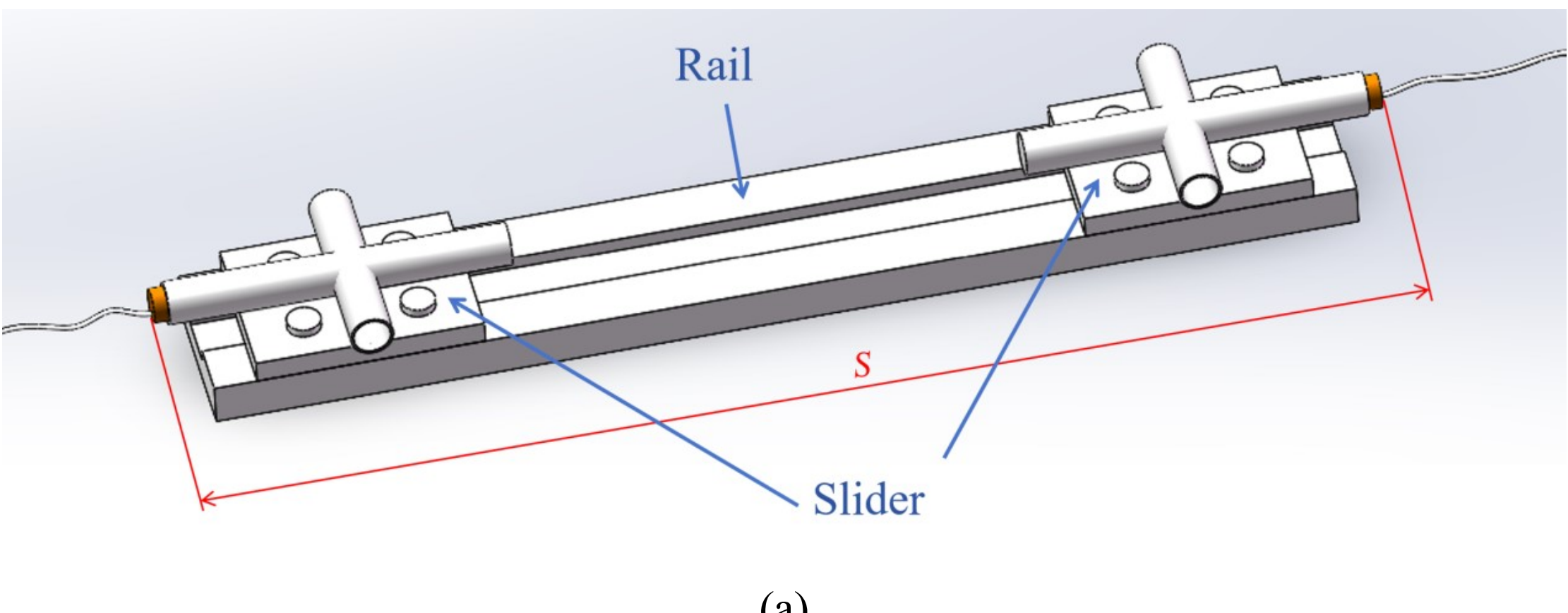

(a)

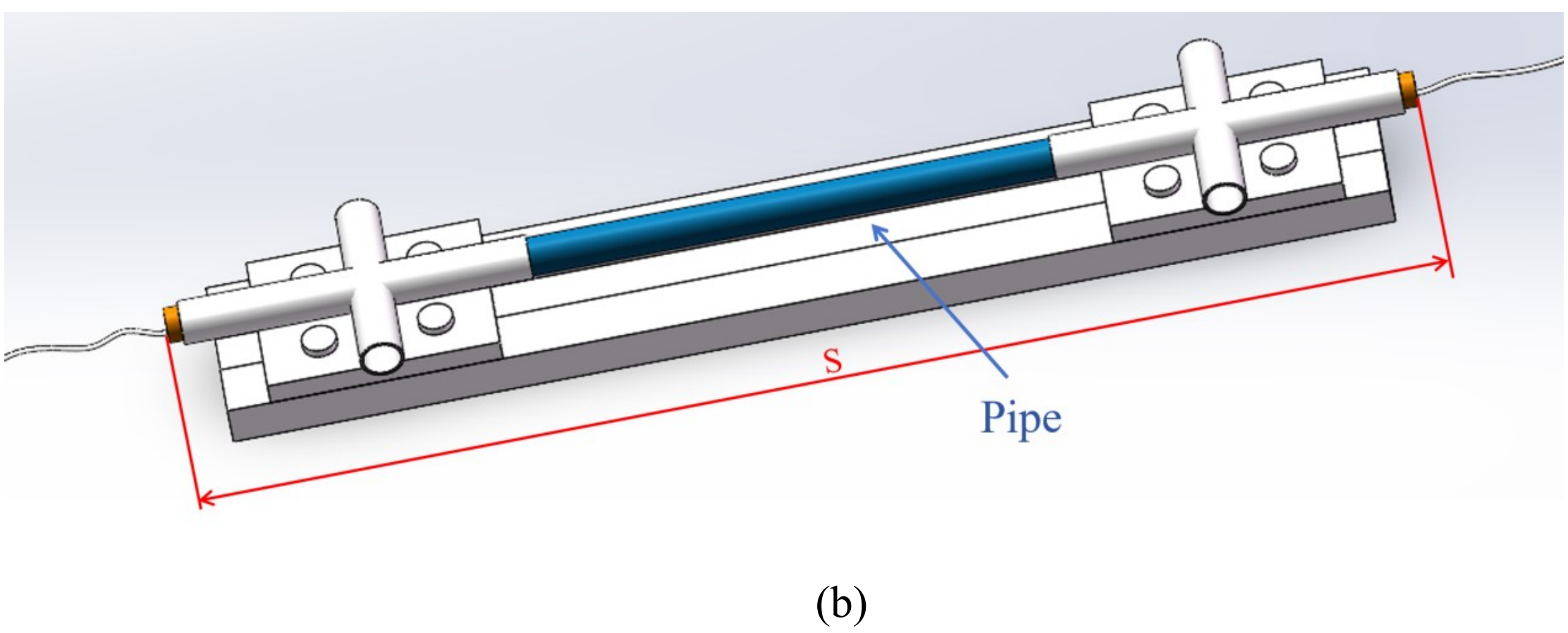

(b)

Figure 11 Experimental setup for measuring the variation of acoustic wave with transducer spacing: acoustic wave propagation (a) in an open environment; (b) inside a pipe.

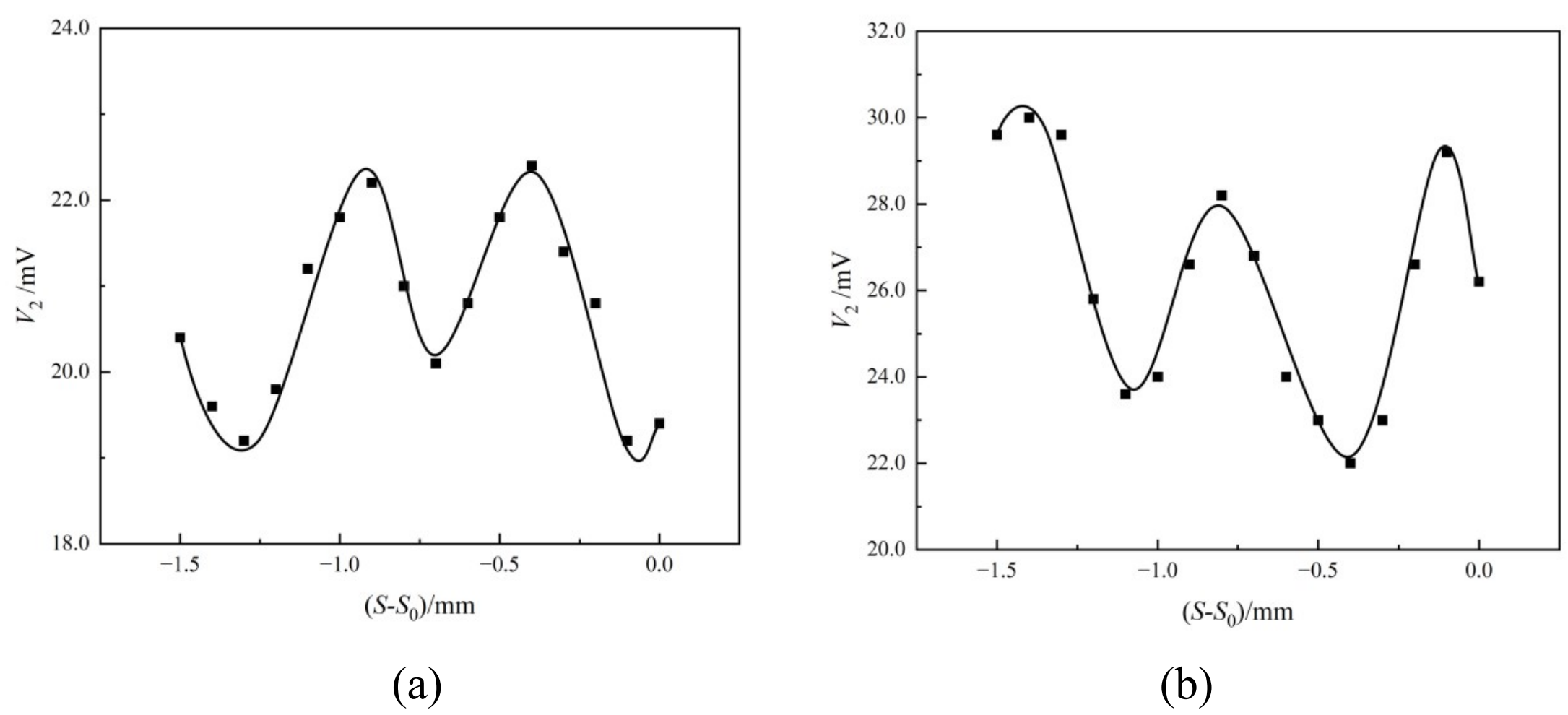

(a) (b)

Figure 12 Variation of the received signal amplitude $V_2$ with transducer spacing $S$ under stationary water conditions: (a) waves in open environment, water temperature 10.2°C; (b) waves inside the pipe, water temperature 8.6°C. Acoustic frequency is $f$=1MHz, the signal voltage of emitter is 2 V and $S_0$=380mm. The standard uncertainty of $V_2$ is approximately 0.1mV.

The results show that in both the open environment and inside the pipe, the received signal $V_2$ exhibits periodic fluctuations as the spacing $S$ changes. According to equation (12), the distance between two adjacent maxima (or minima) of the sound pressure is $\lambda/2 \approx 0.72\text{mm}$. This is approximately consistent with the spacing between adjacent peaks (or troughs) in Figure 12. For the acoustic wave inside the pipe (Figure 12b), the average value of the maxima is 29.1 mV, and the average value of the minima is 22.8 mV. The standing wave ratio is therefore $G = 29.1 / 22.8 \approx 1.276$, and the magnitude of the reflection coefficient is $|r_p| \approx 0.102$ .

### 3.7 Vortex

Turbulence contains vortices of various scales, but its intense random fluctuations indicate a significant distinction from vortices in steady motion. How vortices and turbulence respectively affect acoustic waves?

Previous studies have shown that when acoustic waves encounter vortices during propagation, transmission and scattering occur (Berthet, Fauve & Labbé 2003; Brillant, Chillá & Pinton 2004; Pinton & Brillant 2005). A non-uniform velocity field advects the wave and thus bends the direction of propagation (Lindsay 1948). On the other hand, when the characteristic length scale of the velocity gradients is on the order of the acoustic wavelength, acoustic waves are scattered (Lighthill 1952). However, the aforementioned theoretical and experimental studies primarily focus on fluids that are gases.

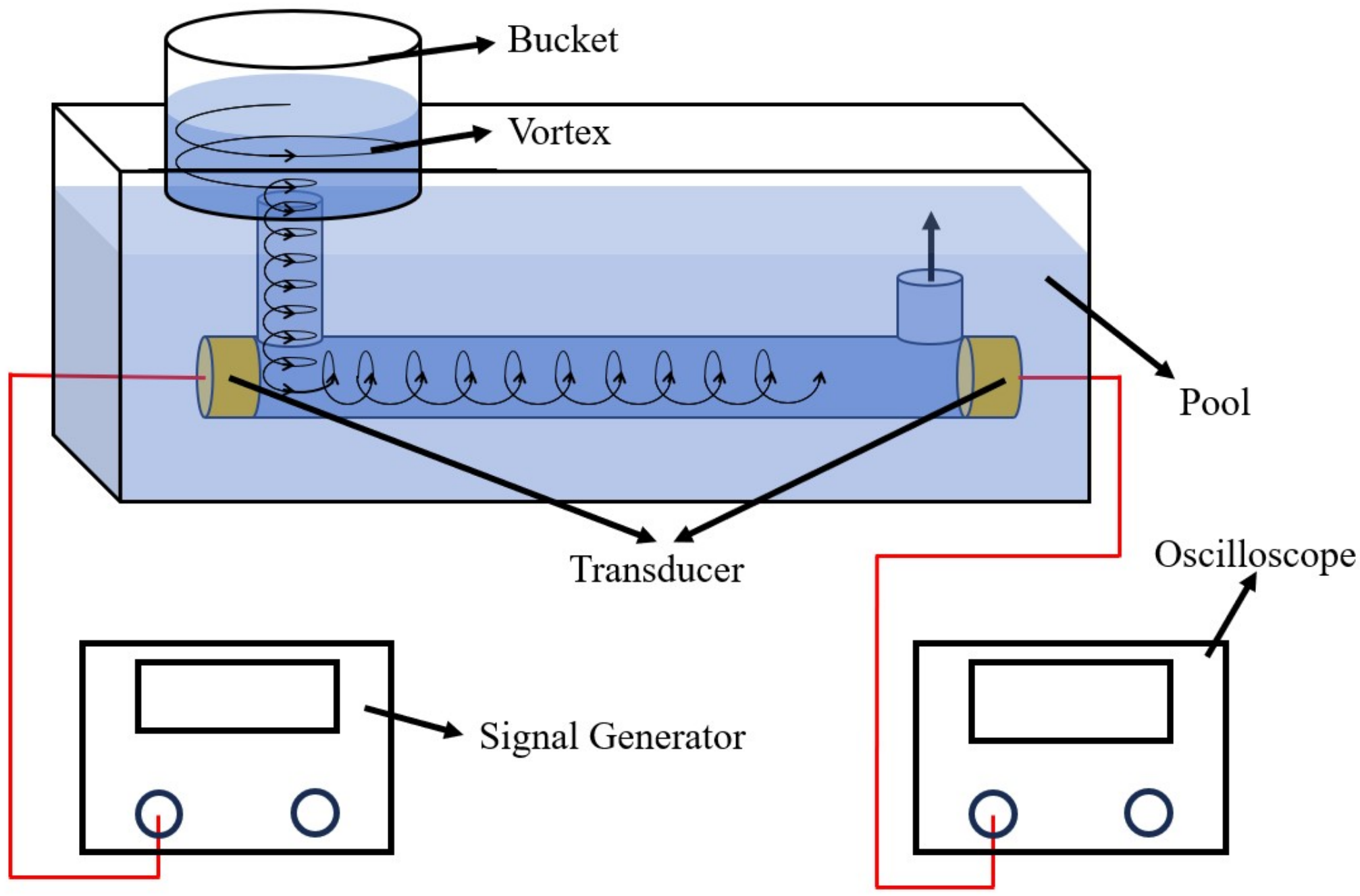

Figure 13　Experimental Setup for the Interaction Between Vortices and Acoustic Waves in Water.

Using the setup shown in Figure 13, the experiments on the effect of vortex in water on acoustic wave propagation are conducted. By initially stirring the water in the bucket, the flow can rotate around the central axis of the bucket. The vortex persists as the water flow downward. Table 4 compares the received signal values when the water is stationary versus when vortex flow is present. It can be observed that there is no difference between the two, indicating that the vortex does not produce a noticeable effect on the acoustic waves (Video 8).

**Table 4**　Voltage signal values for the interaction between underwater vortices and acoustic waves.

| No. | Temperature (°C) | Emitter | | Receiver | |
|---|---|---|---|---|---|
| | | Frequency (MHz) | Voltage $V_1$(V) | Static $V_2$(mV) | Vortex $V_5$(mV) |
| 1 | 12.4 | 0.906 | 5 | 92.8±0.4 | 92.8±0.4 |
| 2 | 13.4 | 1.000 | 1 | 116.0±0.5 | 116.0±0.5 |
| 3 | 12.6 | 1.103 | 2 | 114.5±0.5 | 114.5±0.5 |
| 4 | 12.8 | 1.202 | 2 | 82.8±0.4 | 82.8±0.4 |
| 5 | 13.8 | 1.260 | 15 | 104.0±0.5 | 104.0±0.5 |

To estimate the rotational frequency of the vortex, a dye tracer is added to the water bucket (Video 9). From the video, it can be observed that the dyed filament is approximately at a distance $r_1 \approx 3.8\text{cm}$ from the central axis of the bucket, with a rotational frequency of about 0.5 Hz. The inner radius of the acoustic wave propagation pipe is $r_2 \approx 8.5\text{mm}$. The morphology of the dye lines indicates that the vortex flow remains in a laminar state.

According to the potential flow theory of ideal fluids, the angular velocity at a given point in a vortex is inversely proportional to the square of its distance from the rotational axis. Therefore, the vortex rotational frequency at $r_2$ is approximately 10 Hz, and the frequency closer to the vortex core would be even higher. This order of magnitude is consistent with the turbulent fluctuation frequencies measured in our previous literature (Hu & Hu 2025). However, the difference in acoustic response between the vortex and turbulent fluctuations indicates a fundamental distinction between the two. Turbulent fluctuations inherently possess essential characteristics that differ from those of vortices.

### 3.8 Unsteady laminar flow

A major characteristic of turbulence is its strong fluctuation over time. The following experiment compares the different effects of unsteady laminar flow and turbulence on acoustic waves. In order to generate unsteady laminar flow, one can simply change the water injection at the pipe inlet in Figure 1a to suction, and control the suction duration with a switch to induce unsteady flow within the pipe. Dye line experiments show that the pipe flow generated by suction is laminar (Hu & Hu 2025). When the acoustic waves propagate parallel or perpendicular to the mean flow direction, the received signals show that the unsteady laminar flow has no effect on the acoustic waves.

## 4. Conclusion

Based on our first paper (Hu & Hu 2025), this study conducts research on the impact of turbulence on hydroacoustic waves from more aspects. The changes in phase and amplitude of acoustic waves caused by turbulence at various frequencies are examined.

The Lissajous figures demonstrate that turbulence alters both the amplitude and phase. The total phase shift of the acoustic wave across the entire pipeline equals the sum of the phase shifts of the segments. The amplification factor and phase shift induced by turbulence are independent of the amplitude of the incident wave but are related to its frequency. These phenomena are analogous to the effect of complex refractive index on laser amplification in a gain medium, suggesting that the amplification of acoustic waves by turbulence is likely a process of stimulated emission of in water.

The amplification factor and phase shift by turbulence vary periodically with increasing frequency. The interval between adjacent peaks of the amplification factor, calculated from the resonator filtering theory, shows good agreement with the experimental values obtained in this study. Regarding the phase shift, its variation period differs significantly from that of the amplification factor, and also varies considerably across different frequency bands.

In pipe flow, when the pump is shut off, the turbulence gradually decays thereafter. The amplitude of the received signal undergoes a transient variation process and eventually converges to the value observed under static conditions. The temporal evolution of its amplitude varies with frequency and can be primarily categorized into six distinct types. As for the phase, during the turbulence decay process, the phase shift induced by turbulence monotonically decreases over time.

The experimental results for low-frequency(<7kHz) and high-frequency(>10MHz) acoustic waves show that acoustic waves with frequencies below and above specific thresholds are essentially unaffected by turbulence. When the acoustic wave frequency falls within a specific frequency range, the amplification factor and phase changes induced by turbulence vary continuously with frequency. This phenomenon is very similar to the optical properties of semiconductors.

When two transducers are positioned opposite each other, the incident wave will be reflected on the surface of the receiver. In our experiment, the reflection coefficient is approximately 0.1. The acoustic wave in the flow field can be regarded as a superposition of traveling and standing waves.

Experiments show that unsteady flow and vortices in the laminar state do not produce a noticeable effect on the acoustic waves. Therefore, turbulent fluctuations must possess inherent characteristics that are fundamentally different from both.

## Appendix

The instruments and materials used in this paper, such as the underwater acoustic transducer, signal generator, oscilloscope, linear guide rail, PVC circular pipe, etc., are the same as those in Ref.(Hu & Hu 2025). So the relevant models and parameters are not listed again. The only the new instrument used in this paper is listed below.

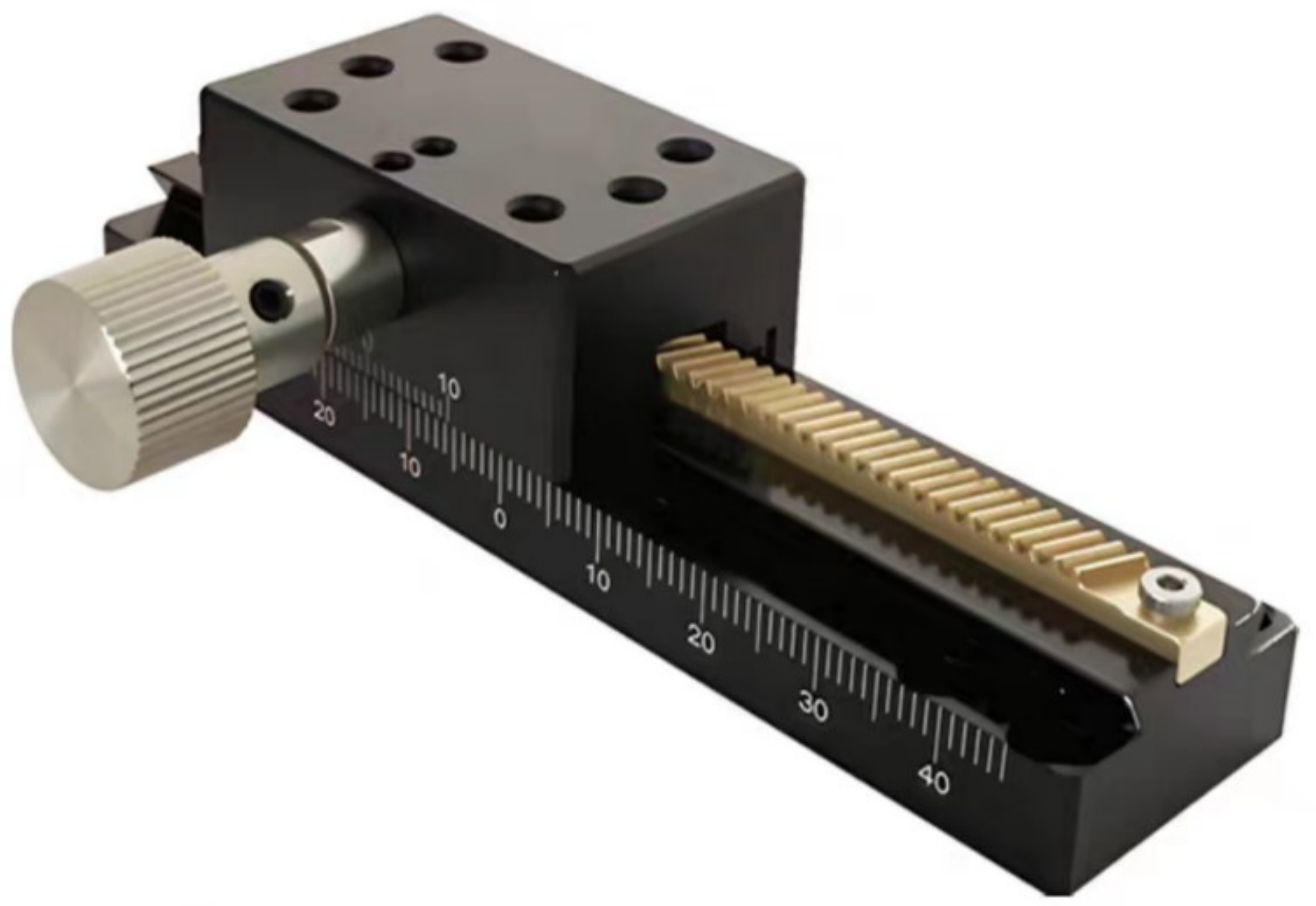

Figure 14 Manual displacement platform.

**Table 5** Instruments

| Instrument name | Model | Manufacturer |
|---|---|---|
| Manual displacement platform | LWX40-100 | Shengshuo Stage |

**Table 6** The parameters in experiments

| No. | Flow | Transducer's fundamental frequency/MHz | Parameters |
|---|---|---|---|
| 1 | Pipe flow | 1,2 | $S$=32cm, $D$=1.7cm, |
| 2 | Pipe flow | 0.2 | $S$=40cm, $D$=7.5cm, |
| 3 | Pipe flow | 0.007 | $S$=71cm, $D$=3.66cm, |
| 4 | Free jet | 1 | $S$=45cm, $D$=1.7cm. |

In the experiments, we observe that the measured values of the acoustic wave reception signals are relatively sensitive to several factors, such as the alignment of the two transducers, the contact condition of the wire connectors, and variations in water temperature and quality. Consequently, the measured amplitude of the acoustic waves may differ after transducers are installed at different times. Therefore, to facilitate quantitative comparisons, all quantitative measurements under similar experimental conditions are conducted within the same time period.

**Data availability**

The authors declare that the data supporting the findings of this study are available within the paper and its supplementary information files. Source data are available upon reasonable request.

## Declaration of Interests

The authors report no conflict of interest.

## Author contributions

Kai-Xin Hu made substantial contributions to the conception of the work, wrote the paper for important intellectual content and approved the final version to be published. He is accountable for all aspects of the work in ensuring that questions related to the accuracy or integrity of any part of the work are appropriately investigated and resolved. Yue-Jin Hu designed and fabricated the experimental apparatus, and conducted experimental measurements jointly with Kai-Xin Hu.

## Acknowledgments

This work has been supported by the National Natural Science Foundation of China (No.12372247), Zhejiang Provincial Natural Science Foundation (No.LZ25A020009), Ningbo Municipality Key Research and Development Program (No. 2022Z213) and the China Manned Space Engineering Application Program—China Space Station Experiment Project (No. TGMTYY1401S).